\documentclass[preprint]{aastex}
\usepackage{epsf}
\usepackage{textcomp}

\begin{document}
\shorttitle{Variability in the NIR disk diagnostic}

\title{Near-IR Variability in young stars in Cygnus OB7}
\author{Thomas S. Rice\altaffilmark{1},
Scott J. Wolk\altaffilmark{2},
Colin Aspin\altaffilmark{3}
}

\begin{abstract}
We present the first results from a 124 night $J$, $H$, $K$ near-infrared monitoring campaign of the dark cloud L 1003 in Cygnus OB7, an active star-forming region. Using 3 seasons of UKIRT observations spanning 1.5 years, we obtained high-quality photometry on 9,200 stars down to $J$=17 mag, with photometric uncertainty better than 0.04 mag. On the basis of near-infrared excesses from disks, we identify 30 pre-main sequence stars, including 24 which are newly discovered. We analyze those stars and find the NIR excesses are significantly variable. 

All 9,200 stars were monitored for photometric variability; among the field star population, $\sim$160 exhibited near-infrared variability (1.7\% of the sample). Of the 30 YSOs (young stellar objects), 28 of them (93\%) are variable at a significant level. 25 of the 30 YSOs have near-infrared excess consistent with simple disk-plus-star classical T Tauri models. Nine of these (36\%) drift in color space over the course of these observations and/or since 2MASS observations such that they cross the boundary defining the NIR excess criteria; effectively, they have a transient near-infrared excess. 
Thus, time-series $JHK$ observations can be used to obtain a more complete sample of disk-bearing stars than single-epoch $JHK$ observations.
About half of the YSOs have color-space variations parallel to either the classical T Tauri star locus 
\citep{Mey97},
or a hybrid track which includes the dust reddening trajectory. This indicates that the NIR variability in YSOs that possess accretion disks arises from a combination of variable extinction and changes in the inner accretion disk: either in accretion rate, central hole size and/or the inclination of the inner disk. While some variability may be due to stellar rotation,  the level of variability on the individual stars can exceed a magnitude. This is a strong empirical suggestion that protoplanetary disks are quite dynamic and exhibit more complex activity on short timescales than is attributable to rotation alone or captured in static disk models.

\end{abstract}


\altaffiltext{1}{Department of Astronomy, Harvard University,
                 60 Garden Street, Cambridge, MA  02138.}
\altaffiltext{2}{Harvard-Smithsonian Center for Astrophysics,
                 60 Garden Street, Cambridge, MA  02138.}
\altaffiltext{3}{Institute for Astronomy, University of Hawaii at Manoa,
                 640 N Aohoku Pl, Hilo, HI  96720.}

\section{Introduction}
\label{data}

Near-infrared studies of young stars allow for the direct detection of optically thick disks around these stars via excess $K$-band flux 
\citep{Lad92, Lad00}.
The intrinsic colors of these disk-bearing young stars occupy a well-defined locus in $(J-H)$ vs. $(H-K)$ two-color space (henceforth called $JHK$ space), and their position along this locus is determined by physical parameters such as their inclination angle, disk inner hole size, and accretion rate 
\citep{Mey97, Rob06}.
This technique of identifying young stellar objects (YSOs) using their near-infrared colors has been used extensively to characterize populations of young stars associated with star-forming regions in Orion, Taurus, Ophiuchus, Chamaeleon, and others for nearly two decades 
 \citep[e.g.][]{Str93,Str95,Ito96,Hil98,Oas99,Hai00,Rob10}.

Disk-bearing young stars, also known as classical T Tauri stars (cTTs), have long been identified as optically variable 
\citep{Joy45,Her62},
with this variability.  At a minimum the variability is due to a combination of (I) cold starspots, (II) hot accretion spots, and (III) circumstellar dust occultations  
\citep{Her94}.
Optical variability can be used to effectively identify young low-mass stars, even in regions lacking other tracers of star formation such as molecular clouds 
\citep{Bri05}.

The variability of T Tauri stars in the near-infrared has been less thoroughly studied. 
One of the first projects was carried out by \citet{Skr96} who sampled bright T-Tauri stars, mostly members of the Taurus cloud, over the course of a few nights.
They found nearly all the stars varied significantly and the amplitude of K-band source variability was weakly correlated with (K$-$L) excess -- a reliable disk diagnostic. From a multi-epoch study of the Serpens cloud core, \citet{Kaas99}  found the IR-variability  a strong indicator of youth.
Variability across 14 epochs was used to reveal new YSOs in the $\rho$ Oph cluster \citep{Alv08}. 
In the early part of the last decade, \citet{Car01}
used 16 observations of a $0\fdg 84 \times 6 \degr$ region over 19 days to study NIR variability toward the Orion Nebula Cluster, 
and 
\citet{Car02}
used 15 observations of a similarly sized 
field in the Chamaeleon I  star-forming region over 4 months;
the typical peak-to-peak amplitude of variability seen was around 0.2 magnitudes in each band.
These studies found  
near-IR (NIR) variability most commonly arose from cold starspots, hot accretion spots, and variable extinction. However, among stars with a near-infrared excess ($\sim 25 \%$ of the variable stars possessed NIR excess), changes in the accretion disk were required to explain the observed variability.
\citet{Eir02} combined optical and NIR data of 18 bright stars and found 12  had correlated optical and NIR variability trends, suggestive of a common physical origin such as spots and/or variable extinction.  The lack of correlation in the other objects was taken as a sign that in those stars, distinct processes were
An investigation of long-term NIR variability in a large sample of T Tauri stars using 2-3 epochs covering baselines of 6-8 years was carried out by 
\citet{Sch09} \citep[see also][]{Sch12}.
They find the fraction of large-amplitude variables increases for progressively longer baselines. They derived a 2500-year upper limit on the duty cycle for large-scale episodic accretion events.  
The YSOVAR survey of an 0.9 $\textrm{deg}^2$ region of the Orion Nebula Cluster 
\citep{Mor11}
obtained 81 epochs of mid-IR Spitzer photometry over 40 consecutive days, in conjunction with 32 epochs of $J$-band images from UKIRT and 11 epochs of $K_s$ photometry from CFHT. In that study of disk-bearing young stars, various phenomena were observed, including periodic variability and disk occultation events.

In this paper, we present a survey in which the photometric variability of objects in the Braid Nebula star-forming region within Cygnus OB7 was monitored for nearly two years. 
Our goal is to detect high-confidence YSO candidates with precision photometry, study their variability, and analyze the stability of the near-infrared disk diagnostic.

The constellation Cygnus contains several rich and complex star-forming regions including Cygnus X as well as the North American and Pelican nebulae 
\citep{Rei08}.
Nine OB associations have been found in Cygnus, with Cygnus OB7 the nearest at a distance of around 800 pc 
\citep[distance modulus $\mu = 9.5$]{Asp09}.
Within Cygnus OB7 lies a complex of several dark clouds collectively known as
Kh 141 
\citep{Kha55}
that have been individually identified in the Lynds catalog 
\citep{Lyn62}.
The dark cloud LDN 1003 in Cyg OB7 has been identified as a site of active star formation, having first been studied in the optical by 
\citet{Coh80}
who found a diffuse red nebula he named RNO 127.
This nebula was later determined to be a bright Herbig-Haro (HH) object by 
\citet[HH 428]{Mel01,Mel03}.
Further study in the optical and near-infrared identified a number of Herbig-Haro objects 
\citep{Dev97,Mov03}
and multiple IRAS sources 
\citep{Dob96}
that reveal the presence of a young stellar population and significant star formation activity.


The presence of two FUors in this region \citep{Rei97,Gre08,Asp11} that has come to be known as the Braid Nebula region \citep{Mov06} led to a focused, multi-wavelength effort to study the young stars in this dark cloud, which is currently ongoing. 
A narrowband optical and near-infrared survey of HH outflows observed using the Subaru 8 m telescope 
\citep{Mag10}
found 12 outflows that have identifiable originating/exciting sources and many more nebulous objects not yet associated with identified sources.
\citet{Asp09}
carried out a near-infrared integral field spectroscopic survey to identify actively accreting sources such as T Tauri stars,
which studied 16 sources in the field and identified 12 young stars among them.
Using the Caltech Submillimeter Observatory, a 1.1 mm map of cold gas and dust associated with these young stars was obtained
\citep{Asp11},
identifying 55 cold dust clumps, 11 of which were associated with IRAS sources.
The wide range of evolutionary states encountered in this region, from starless clumps to optically visible T Tauri stars, suggests that star formation here is an ongoing process rather than a one-time occurrence. 

In this paper we describe a $JHK$ monitoring survey of the Braid Nebula star-forming region in Cygnus OB7. 
We plan to present first results from this survey in two papers; this is paper 1 of 2. 
Our goal in Paper I is to identify disk-bearing young stars,
to broadly characterize the variability properties of these stars using sensitive, long-baseline, nightly-cadence time-series photometry to investigate the reliability of the near-infrared excess criterion itself with respect to time.
In the second paper (Wolk et al. 2012, in prep, hereafter known as Paper II) we will describe the specific phenomena seen in the lightcurves, carefully analyze the variability and periodic nature of specific stars and present variability statistics for the field star population.


In \S \ref{data} we outline the UKIRT observations and data reduction procedures.
In \S \ref{disks} we describe our method of identifying stars with near-infrared excesses and studying their variability.
In \S \ref{results} our results are presented, 
and in \S \ref{discussion} we discuss the broad implication of these results on infrared studies of young stars.

\section{Data}
\label{data}

\subsection{Observations and Data Processing}
\label{observations}

\begin{figure}
\plotone{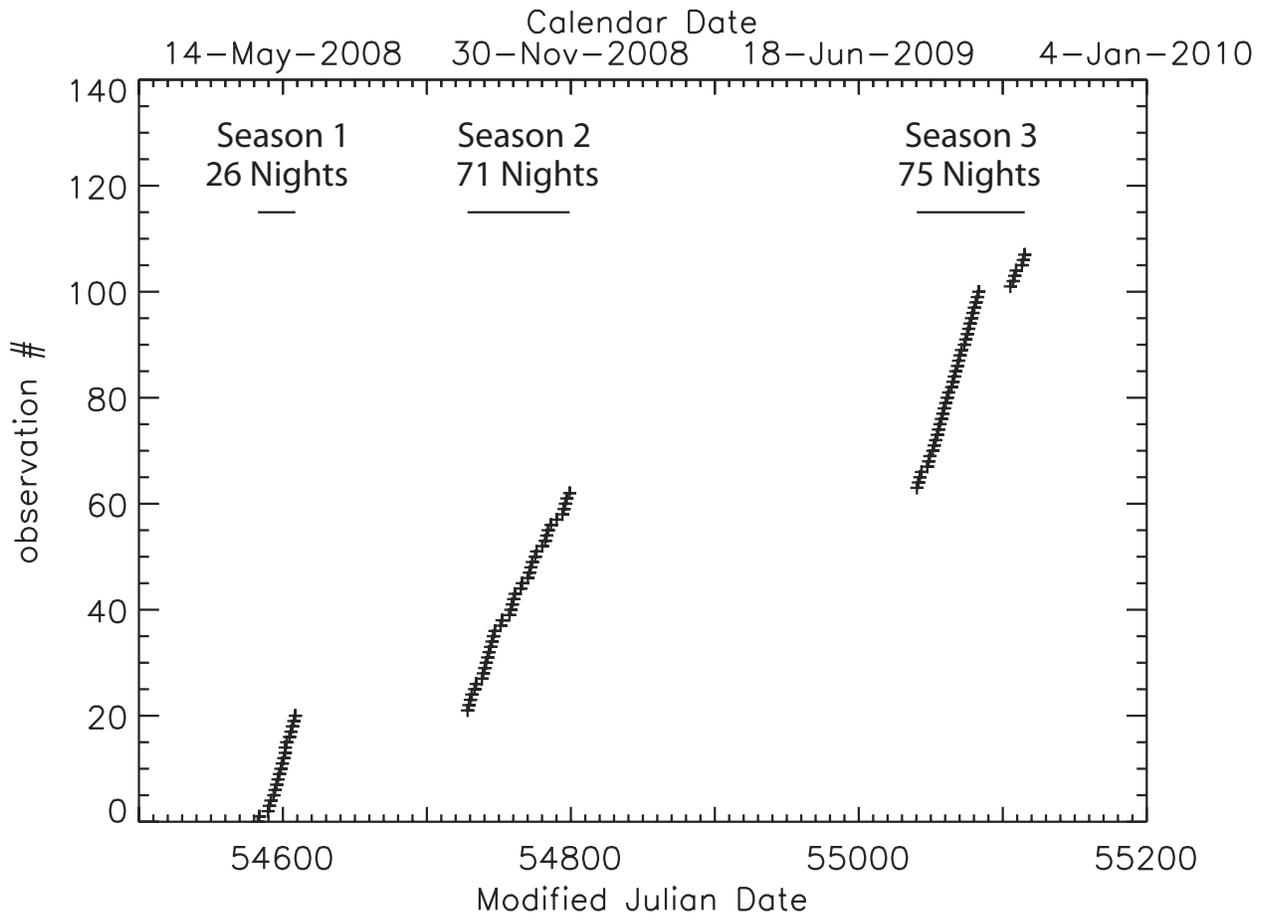}
  \caption{Schematic representation of when data were taken for this study.  Nights with high errors have been removed (See \S 2.2).}      
    \label{fig:obs.log}
\end{figure}

The $J$, $H$, and $K$ observations of a $0.9^\circ \times 0.9^\circ$ region in Cygnus OB7 were obtained using the Wide Field Camera (WFCAM) instrument on the United Kingdom InfraRed Telescope (UKIRT), an infrared-optimized 3.8 m telescope atop Mauna Kea, Hawaii at 13,800 feet elevation. 
These data consist of WFCAM observations taken from 26-April-2008 to 11-October-2009 (Figure~\ref{fig:obs.log}) in three observing seasons as part of a special observation program described in  Aspin et al. 2012 (in preparation). The first season was in spring of 2008 and covered 26 nights. The second season was during fall 2008 and lasted 71 days. The third season was approximately one year later and lasted 75 days.  During the monitoring runs, data were taken once per night on a total of 124 nights in all three bands.  Atmospheric seeing was between $0\farcs5$ and $1\farcs6$ on any given night. 
The $J$, $H$, $K$ filter bands on WFCAM comply with the $JHK$ broadband filters from the Mauna Kea Observatories Near-Infrared filter set 
\citep{Cas07,Tok02};
WFCAM's photometric system is described in 
\citet{Hew06}.

\begin{figure}
  \plotone{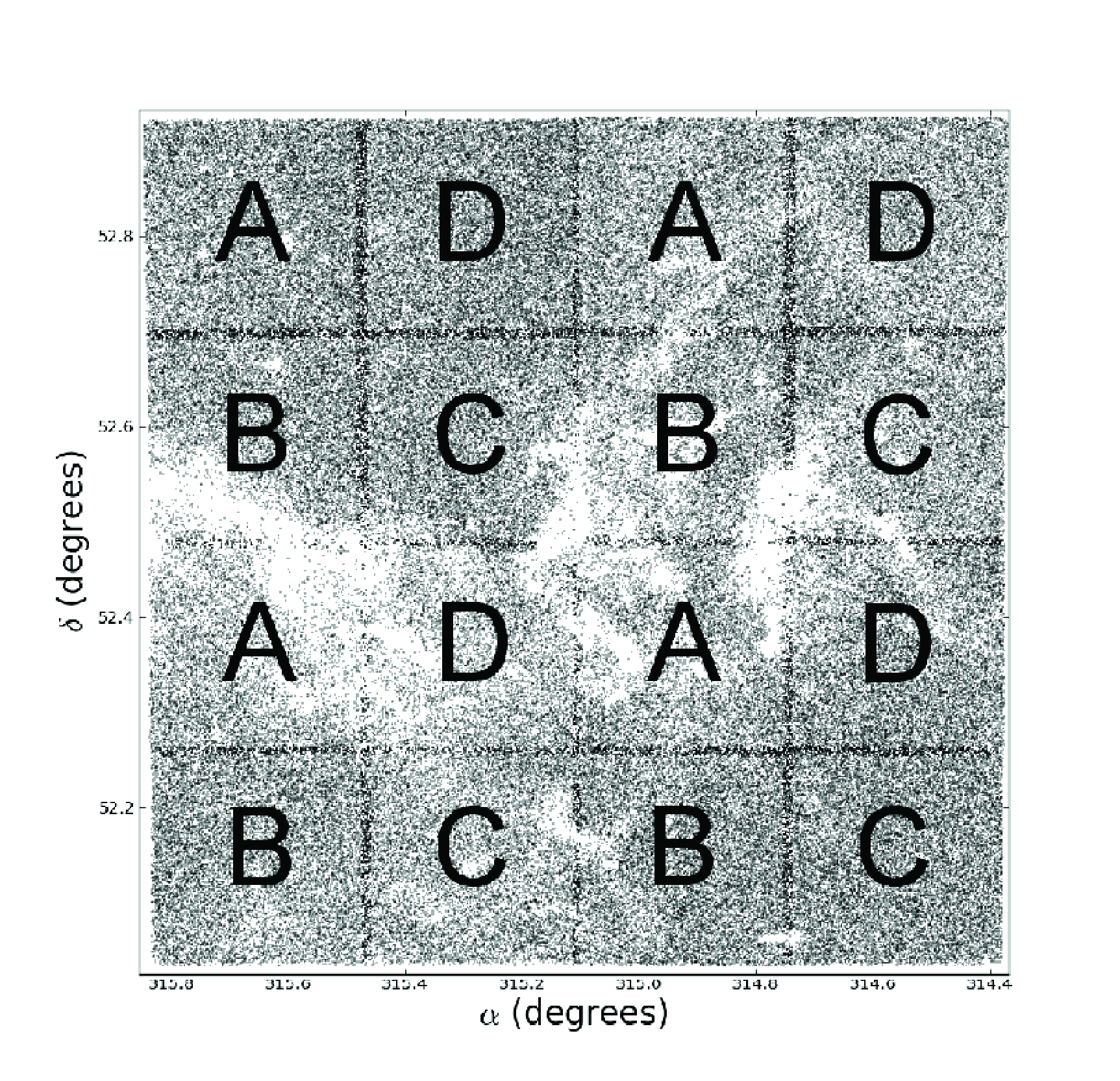}
  \caption{
    The footprint of WFCAM consists of four detectors spaced by (94\%) of
    their width, covering a non-contiguous 0.21 square degrees; four pointings
    are required to fill the observing field. 
    In this figure we show how the 16 tiles in this observing field are imaged:
    the 4 tiles marked ``A'', called ``footprint A'', are observed in a
    simultaneous imaging, then the telescope is slewed south by $15\arcmin$ to
    observe the ``B'' tiles, west by $15\arcmin$ to observe ``C'' tiles, and
    finally back north $15\arcmin$ to observe the ``D'' tiles. 
    Underlaid is a map of star counts across the field, showing clearly the
    structure of the dark cloud L 1003, which causes the mean extinction in
    each tile to vary.
    \label{fig:wfcam}
}
\end{figure}

Detailed specifications for the WFCAM instrument are given in 
\citet{Cas07}.
The instrument has four 2048 $\times$ 2048 Rockwell Hawaii-II PACE arrays with a scale of 0.4 arcseconds per pixel, giving a combined solid angle of 0.21 square degrees per exposure. 
The four detectors are widely spaced, at 94\% of one detector's width apart 
(see Fig. \ref{fig:wfcam}). 
We used a standard effective integration time of 40 seconds per pointing. 
In a stepping pattern, WFCAM can scan a nearly-complete square degree of the sky in four pointings 
(A, B, C, D in Fig. \ref{fig:wfcam}). 
The pointings have a minor overlap on their edges such that stars in the overlapping region were observed twice per night.

Data from the survey were pipeline-reduced and processed by the WFCAM Science Archive System 
\citep{Irw08,Ham08},
which is also used for the UKIDSS survey described in 
\citet{Law07}.
Near-infrared observations are calibrated against 2MASS sources in the field which have 
extinction-corrected color $0.0 \le J - K \le 1.0 $ and  2MASS signal-to-noise ratio $> 10$ in each filter.
For the target stars, total uncertainties in photometry are typically 2\% down to $J$=16.5, $H$=16, and $K$=15, and errors are less than 4\% at $J$=17 
\citep{Hod09}.
In processing such a wide field of view, a large number of data quality issues arise and are typically dealt with by the pipeline by assigning photometric error flags for issues such as bad pixels, deblending, saturation, and other effects.

\subsection{Data Retrieval \& Cleaning}
\label{cleaning}

We retrieved the processed photometry data from the WFCAM Science Archive website via an SQL interface. 
All 124 nights of data were retrieved, cross-matched, and merged together into a single catalog containing columns for object ID, observation date, sky coordinates, $JHK$ photometry, and various photometric processing flags. 
Our initial query was for data that satisfied the criteria $J$, $H$, $K < 18$; $J$, $H$, $K > 9$; and photometric uncertainty $\sigma_{(J,H,K)} < 0.1$, in order to include all possibly relevant data on young stars in this region 
(see \S \ref{completeness} on magnitude cuts) 
while keeping the downloaded catalog file at a manageable size. 

\citet{Hod09}
present an empirically derived correction from the pipeline-estimated photometric error to the true, measured error:
\begin{equation}
  M^2=cE^2+s^2  
\end{equation}

\noindent where $M$ is the measured total error, $E$ is the estimated photometric uncertainty given by the pipeline, the constant of proportionality $c=1.082$, and the systematic component $s=0.021$. 
We applied this update to the estimated photometric uncertainties after retrieving the data and confirming that night-to-night variations at the 2\% level were typical even for high signal-to-noise stars.

\begin{figure}
  \plotone{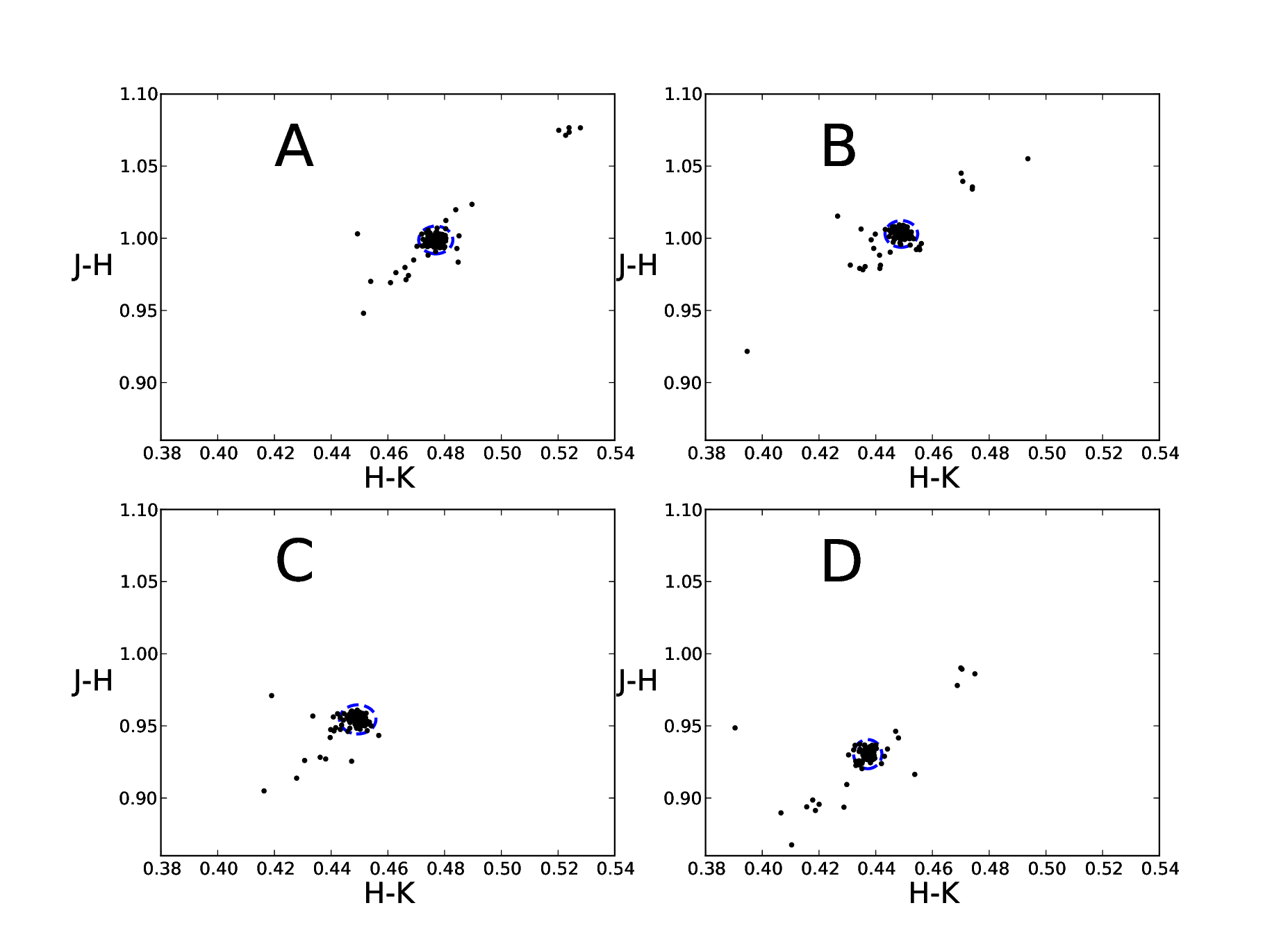}
  \caption{
    An illustration of our procedure to reject suspicious observations.
    On each night, we calculated the mean color of a large sample 
    of stars in each footprint. 
    Each nightly mean color is plotted as one point 
    for footprints A, B, C, and D 
    (see Fig. \ref{fig:wfcam}). 
    The dashed blue ellipses enclose the nightly 
    mean colors for observations considered ``reliable''. 
    Nights with mean colors outside that region were rejected.
    \label{fig:reject}
  }
\end{figure}

To assess the photometric integrity of this dataset, we calculated the mean $JHK$ colors for each pointing on each night, averaged over all stars. 
We excluded observations where the mean colors showed significant deviations. 
For each night, we computed the mean $J-H$ and $H-K$ color of all reliable stars within each detector footprint 
(here, ``reliable''  denotes stars with photometric uncertainties less than 0.1 magnitude in each band and no processing error flags, while avoiding bright stars). 
In practice, this translated to stars between 
$13<J<18$, $12<H<17$, and $11.5<K<16.5$. 
Typically, 25,000 stars met this criterion every night. 
We find that systematic night-to-night color deviations of the ensemble on each footprint are indeed about two percent (as expected), but a significant minority ($\sim 15\%$) of nights exhibit large offsets in color space 
(see Fig. \ref{fig:reject}), 
likely due to non-uniform extinction from thin clouds. 
We applied a form of iterative outlier clipping to select and remove anomalous nights from our analysis, leaving only the nights whose mean colors lay within $3 \sigma$ of the outlier-clipped, time-averaged global mean color.

We also note that the mean color in each footprint is significantly different. 
This is because each footprint samples regions of different visual extinction, as seen in Fig. \ref{fig:wfcam}. 
Instrumental effects are not expected to cause this, as all 4 detectors are included in each footprint.

Out of the 124 nights in the original survey, 24 nights were rejected due to significant deviations in the mean color, leaving 100 nights for our analysis (Fig.~\ref{fig:obs.log}).

\section{Disk Identification and Analysis}
\label{disks}

Our scientific goal in this project is to detect young stars that possess disks by their $K$-band excess, to briefly characterize the variability of these disked stars, and to investigate the stability of these $K$-band excesses with respect to time. 

\subsection{The Near-Infrared Excess}
\label{excess}

To detect optically thick disks around young stars, we use the near-infrared excess criterion developed by 
\citet{Lad92}
 and the classical T Tauri star (CTTS) locus reported by 
\citet{Mey97}.
We consider stars to have a near-infrared excess consistent with an optically thick disk at $K$-band 
(hereafter referred to simply as a ``$K$-band excess'')
if their colors fall significantly to the right of $JHK$ space demarcated by the main sequence reddening band and within the CTTS locus. 
To ensure the disk signatures are significant, 
we require the sources to be 4$\sigma$ red--ward of the reddening vector associated with the reddest, non-disk bearing stars \citep{Lad92}:
\begin{equation}
  (J-H) \le 1.714 \times (H-K) 
\end{equation}

\noindent and not more than 4 $\sigma$ below an empirically derived locus of the de-reddened location of about 30 CTTS on the near IR color-color diagram \citep{Mey97}:

\begin{equation}
  (J-H) \ge 0.58 \times (H-K) + 0.52
\end{equation}   

\begin{figure}
  \plotone{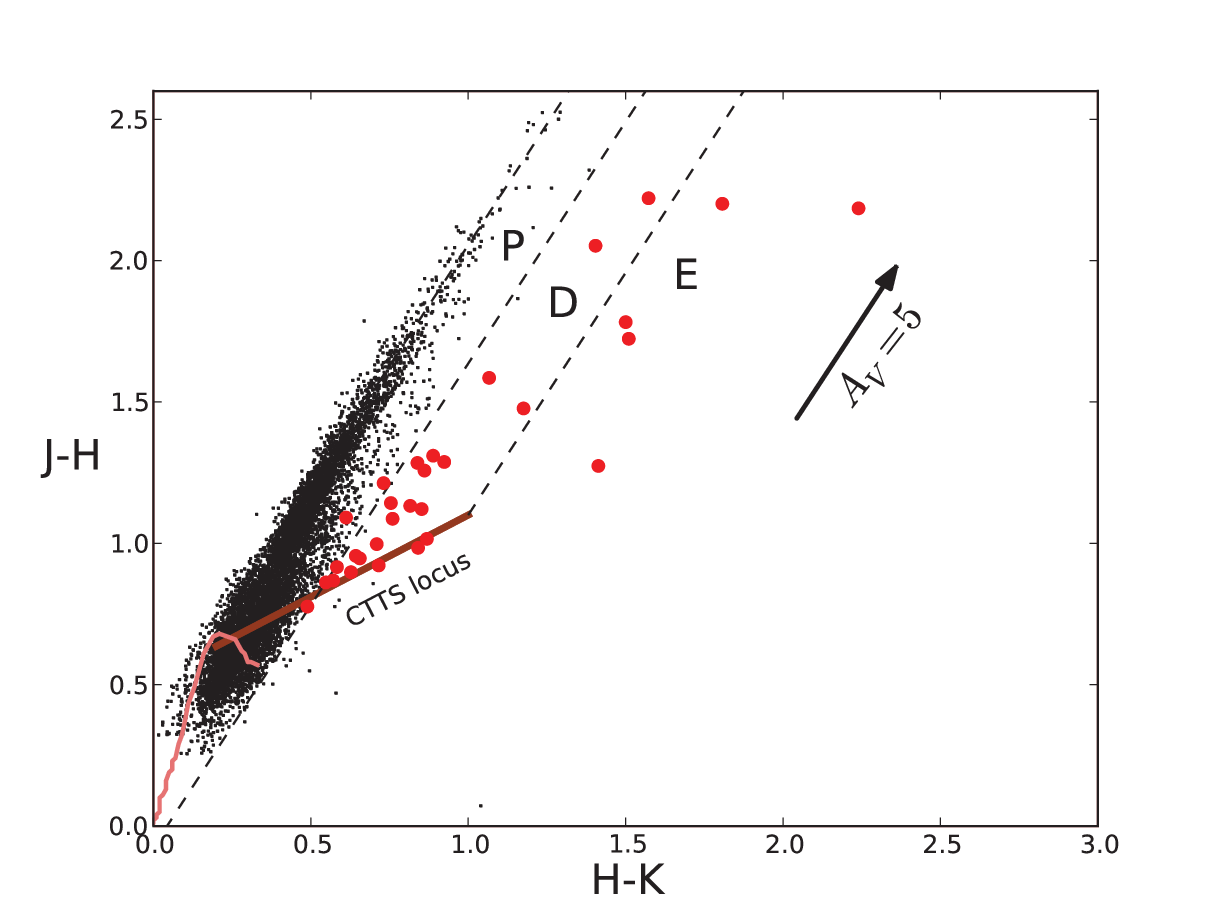} 
  \caption{
    A JHK color-color diagram showing the mean colors of the 9200 stars
    included in our analysis plotted as black points. 
    The meaning of regions ``P'', ``D'', and ``E'' are explained 
    in the text at \S \ref{excess}. 
    The thirty stars that we identify as disked are plotted as 
    red circles; 
    three stars (10\%) lie, on average, in region ``P'', but 
    are considered disk-bearing due to their observed variability 
    that moves them into region ``D'' (see \S \ref{transient}).
    The solid line is the locus of main-sequence stars 
    \citep{Koo83}
    and the CTTS locus 
    \citep{Mey97}.
    Dashed lines are parallel to the reddening vector 
    \citep{Rie85},
    and a reference reddening vector corresponding to 5 magnitudes
    of visual extinction ($A_V = 5$) is shown as a solid arrow. 
    \label{fig:color}
  }

\end{figure}

We show the distribution of stars within $JHK$ space in Figure \ref{fig:color}. Underplotted is the locus of main-sequence star intrinsic colors 
\citep{Koo83}
as a solid curved line, 
and the classical T Tauri star locus 
\citep{Mey97}
as a solid straight line that terminates near (1.0, 1.0) in $JHK$ space,
corresponding to the highest accretion rates and smallest disk hole sizes found by
\citet{Mey97}.
Reddening vectors using the extinction law presented in
\citet{Rie85}
are plotted out from the tip of the CTTS locus and the main-sequence curve.
Loosely following 
\citet{Ito96},
we partition the inhabited areas of $JHK$ space into 3 regions: ``P'', ``D'', and ``E'', meaning ``photosphere'', ``disk'', and ``extreme'' respectively, demarcated by these reddening vectors.

Region ``P'' is inhabited by stars whose NIR emission is dominated by their photosphere. 
This includes main sequence stars, giants, and some pre-main sequence stars with a small or negligible $K$-band excess, including CTTS with small $K$-band excess, as well as weak T Tauri stars (wTTs). 
Single-epoch or time-averaged near-infrared colors cannot distinguish between main sequence stars and YSOs that lie in this region.

Region ``D'' is occupied by stars whose NIR emission originates from both a photosphere and a disk, and is consistent with simple models of an accreting, optically thick disk at $K$-band 
\citep{Mey97}.
All stars in ``D'' are definite disk-bearing young stars, but disked stars can also occupy Regions ``P'' or ``E'', so stars in Region ``D'' are not a complete sample of disk-bearing stars.

Region ``E'' contains stars with more excess at $K$-band than can be accounted for by an accreting, geometrically flat disk. These stars will be hereafter referred to as ``extreme $K$-excess stars'', and are expected to be less-evolved; their redder colors may be due to emission from a circumstellar envelope. Class I protostars have been found to inhabit this region due to their redder colors 
\citep{Lad92, Rob06}.

\subsection{Study Depth}
\label{completeness}

Our goal is to search for high-confidence pre-main sequence stars in Cygnus OB7. 
We chose a J=17 brightness cut. This limits errors in $J$ to about 4\% with similar errors in $H$ and $K$ for typical stellar colors. 
This reduced our input catalog to 9,200 stars. 
At the published distance of Cyg OB7 (800 pc; distance modulus $\mu=9.5$),
we can estimate to what stellar mass depth this survey reaches by using pre-main sequence isochrones calculated from 
\citet{Sie00}.
For these isochrones we assume a typical YSO age of $10^6$ yr.
The most extinguished YSO in this sample is seen through about 11.5 magnitudes of visual extinction, as estimated by tracing its $JHK$ color back to the CTTS locus and measuring the resulting color offset in units of $A_V$.
Assuming a maximum extinction of $A_V = 12$, this survey is reaches a nominal depth of  0.3 $M_\sun$, and in less extinguished regions where $A_V < 7$ we  should reach down to the hydrogen-burning limit ($\sim 0.1 M_\sun$).  However, since the deepest part of the clouds have not been penetrated by the survey we have no knowledge of the maximum extinction, nor any depth to which we can be assured we are complete. 



\subsection{Variability}
\label{variability}

We identify a star as ``variable'' if it is seen to change at a level greater than its photometric noise. To quantitatively select stars that are variable in this dataset, we use the Stetson variability index $S$ 
\citep{Ste96, Car01}.
The Stetson index is useful for multi-wavelength simultaneous observations, as it assumes that true variability will cause observations at different wavelengths to rise or fall in unison;
its usefulness as a criterion for variability has been established by multiple time-series studies
\citep[e.g.][]{Car01,Car02,Pla08,Mor11}.
The Stetson index identifies variables even among stars whose variability is comparable to photometric noise without any assumptions about the type of variability seen, except that true variability should cause all channels to vary.

The Stetson index is computed by the following equation:


\begin{equation}
  S = { \sum_{i=1}^{p} 
  \textrm{sgn} \left(P_i\right)\sqrt{\left|P_i\right|}}
\end{equation}

\noindent where $p$ is the number of pairs of simultaneous observations of a star.
$P_i = \delta_{j(i)}\delta_{k(i)}$ is the product of the relative error of two observations.
The relative error is defined
as:
\begin{equation}
  \delta_i = \sqrt{\frac{n}{n-1}}\frac{m_i - \bar{m}}{\sigma_i}
\end{equation}

\noindent for a given band. The size of the bias is $\sqrt{{(n-1)}/{n}}$ where $n$ is the total number of observations contributing to the mean. The second term is the standard error term, where $m_i$ is the measure magnitude, $\bar{m}$ the mean magnitude and $\sigma_i$ the intrinsic error of the individual measurement.  

\begin{figure}
\plotone{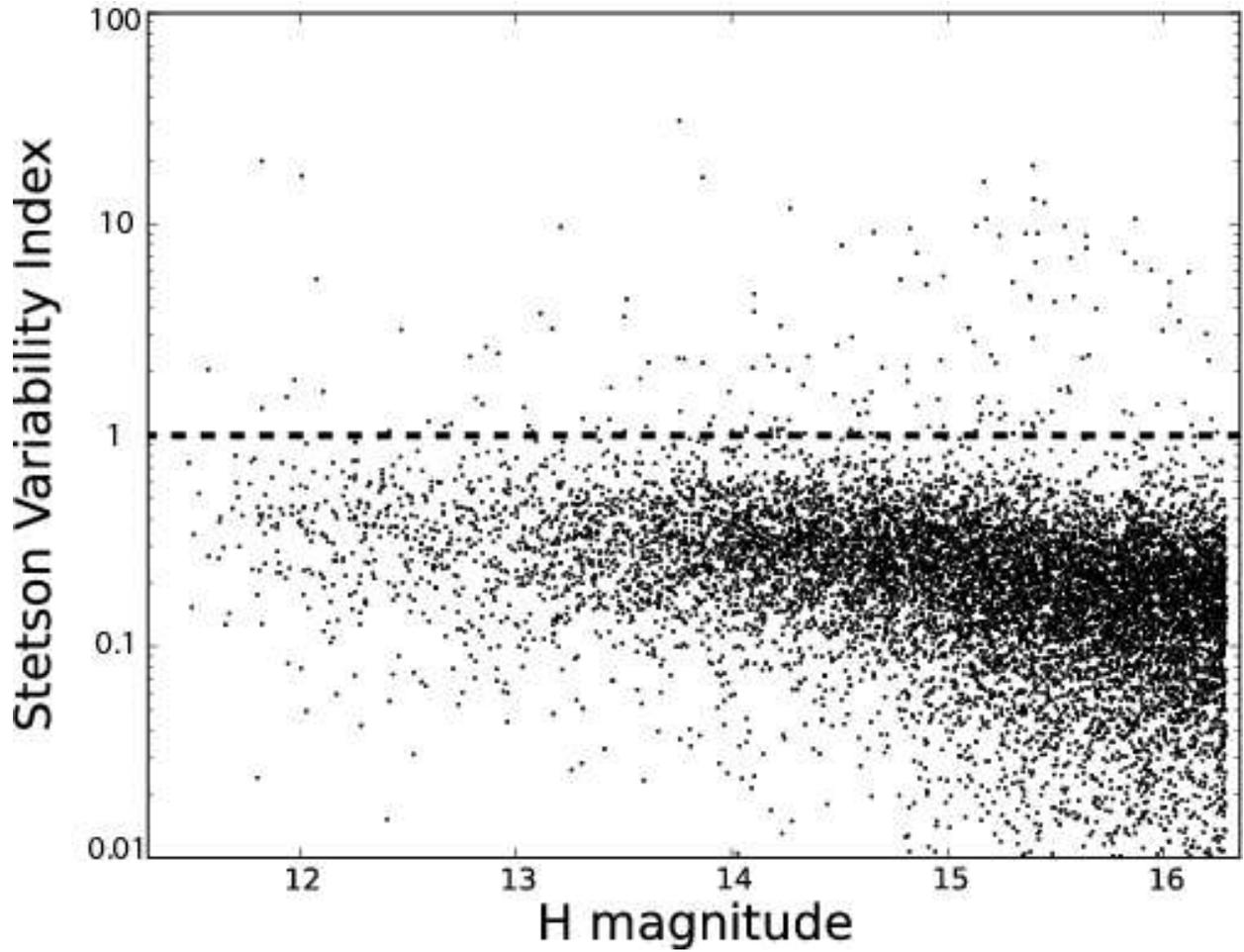}
  \caption{The value of the Stetson index for all 9200 stars. As a function of H magnitude the distribution is flat with a typical value of about 0.2.  The threshold of 1 is about a 4 $\sigma$ deviation and is exceeded by $\sim$160 sources.
        \label{stetson}
  }

\end{figure}

Formally, the Stetson index is designed to identify stars as variable when $S > 1$ if photometric uncertainties are properly estimated. After applying the error correction described in \S \ref{cleaning}
and calculating $S$ for all 9,200 stars, we find the outlier-clipped mean $S$ value to be 0.2, with the outlier-clipped distribution having a standard deviation of 0.16. Therefore, stars with $S \ge 1$ can be considered $5\sigma$ variables, and we use $S \ge 1$ as our criterion for variability (Fig~\ref{stetson}).

\begin{figure}
  \plotone{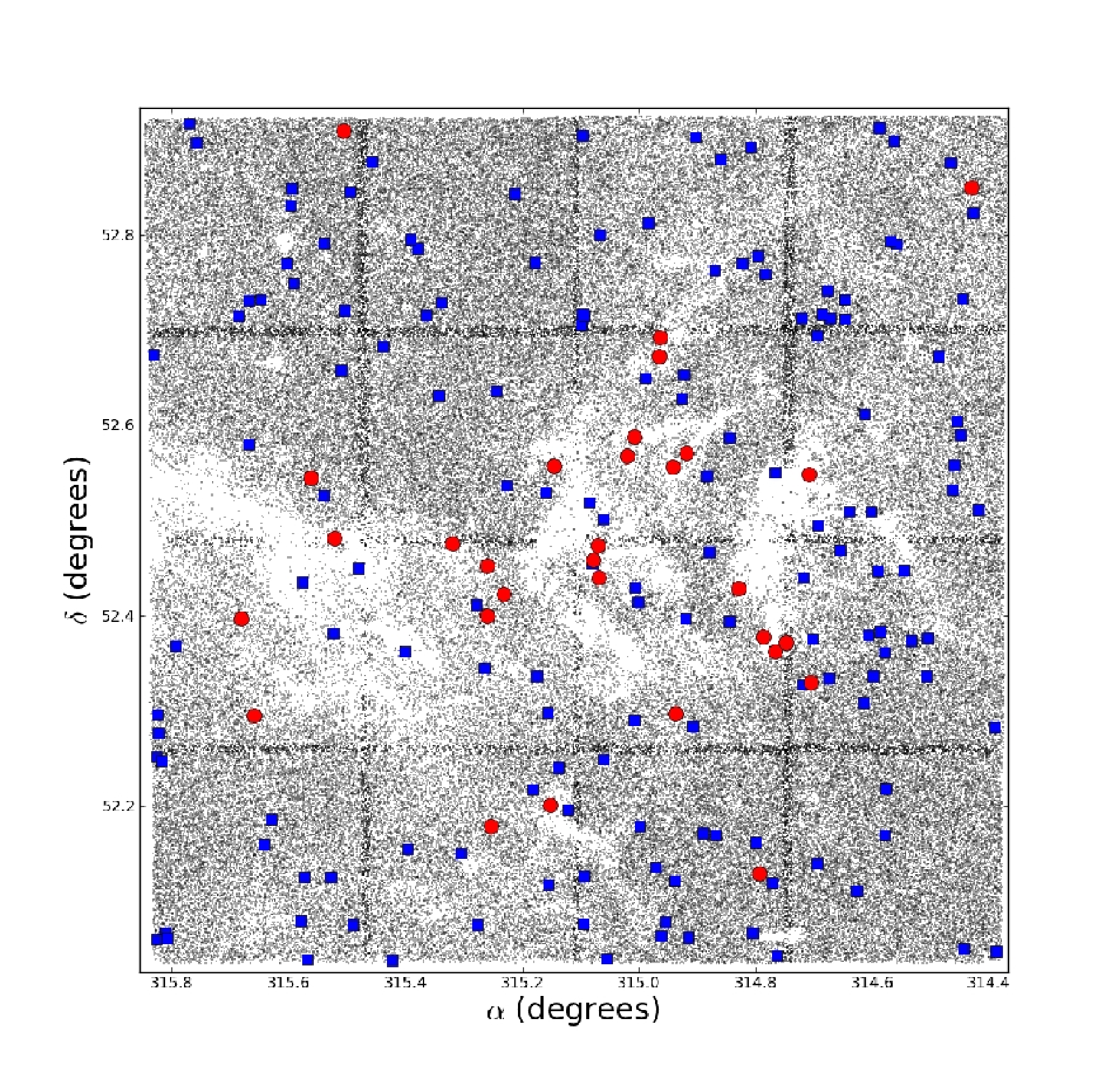}
  \caption{
    The spatial distribution of disked and variable stars detected 
    in our analysis. Disked stars are plotted as red circles; 
    variable stars that lack $K$-band excess are plotted as blue squares. 
    Most (90\%) disked stars lie within the boundaries of the dark cloud,
    while variables are found uniformly in the field. 
    \label{fig:map}
  }
\end{figure}

All stars brighter than $J$=17 with no photometric processing error flags were analyzed for variability. Among these 9,200 field stars, we recover $\sim $160 that are variable according to the Stetson index $S > 1$. 
The positions of these 160 stars are plotted in Fig. \ref{fig:map} as blue squares. 
In this paper, we focus on the identification and variability characteristics of the disked population; the variability characteristics seen in the non-disked stars will be discussed in Paper II.

\subsection{Transient Excesses}
\label{transient}

If a star exhibits a $K$-band excess in only a fraction of its observations, we consider its $K$-band excess to be transient.
We do not expect the circumstellar disks of such stars to actually disappear and reappear; rather, the disks in such systems are likely undergoing physical changes that cause their $H-K$ colors to vary back and forth across the line demarcating unambiguous disked stars (region ``D'') from ambiguous main sequence stars (region ``P''). Such a change could feasibly be induced by, star spots (hot or cool), impulsive heating events such as stellar flares, changes in (inner) disk inclination, local extinction, central hole size, or a varying accretion rate
\citep{Bou89, Mey97, Sch12}.


To identify and characterize stars with transient $K$-band excess, 
all data satisfying our quality filter 
were evaluated against 
Equations 2 and 3 (see \S \ref{excess}). 
Stars that showed a $K$-band excess according to these criteria were tallied, producing a table of stars with $K$-band excess in at least one observation, along with the number of times that star was observed and the fraction of nights that the star displayed a $K$-band excess. 

\begin{figure}
  \plotone{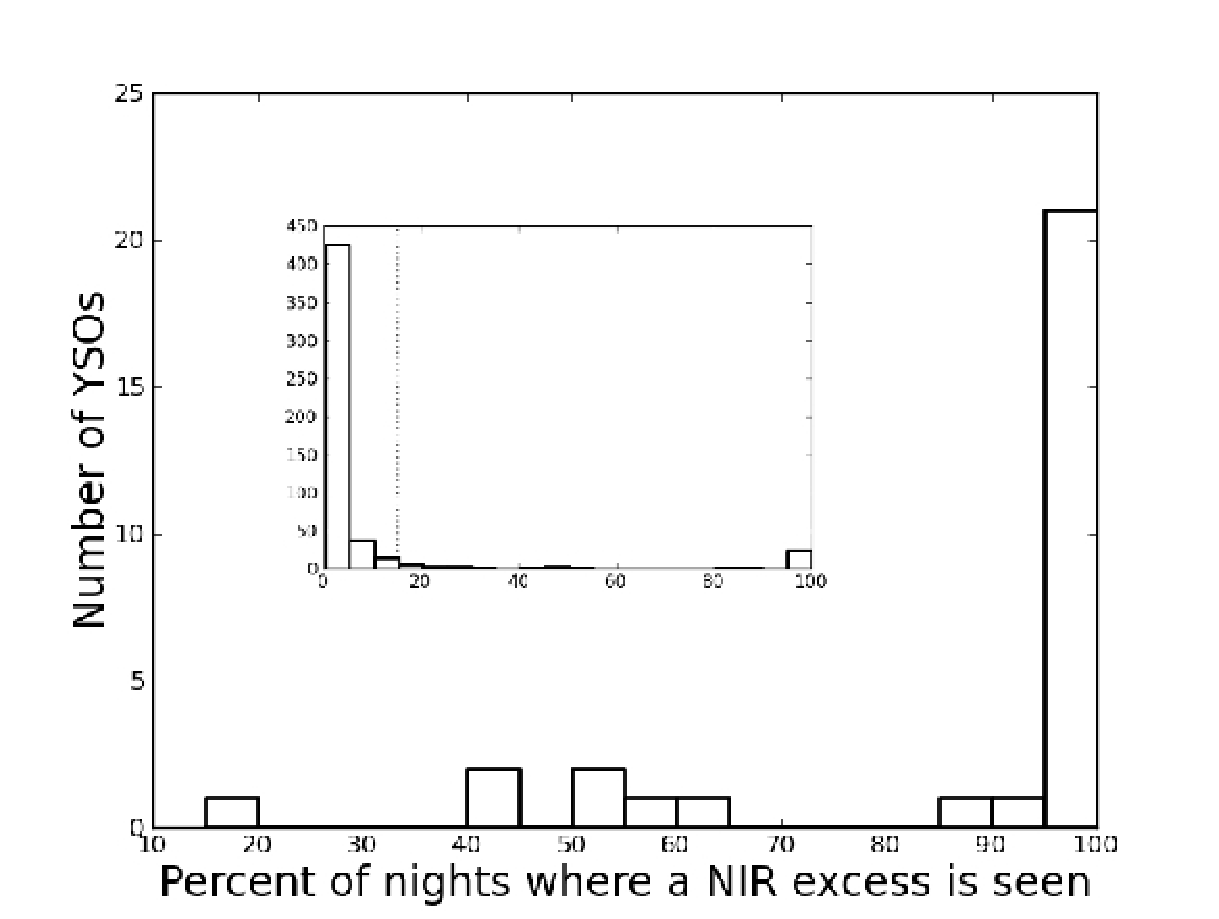}
  \caption{
    A histogram showing how steady the near-infrared 
    excess was in each of the 30 YSOs. 
    Inset is the raw sample showing the 528 stars with a 
    $K$-band excess on at least one night. All stars with 
    $K$ excess on less than 15\% of nights were rejected as 
    false positives (dotted line).
    Most of the confirmed YSOs show a consistent NIR excess 
    on every night or nearly every night, but a significant 
    minority of the YSO sample (seven stars) exhibit a transient 
    $K$-band excess.  
    \label{fig:hist}
}
\end{figure}

We find 528 stars that show a $K$-band excess on at least one night. 
Given our $4\sigma$ cutoff, we expect a substantial number of single-night false positives due to photometric noise assuming Gaussian statistics in $\sim 920,000$ individual observations. We filter most of these false positives by removing all stars that show a near-infrared excess in fewer than 15\% of nights or those that  met our measurement criteria on 25 or fewer nights. 
(See Figure \ref{fig:hist}, inset.)
This cut makes us insensitive to any YSOs who genuinely possess a disk that contributes to a significant $K$-band excess in less than 15\% of observations, but it filters out virtually all false positives while allowing us to remain sensitive to stars with small and moderate, but stable, $K$-band excesses.

These criteria identify 42 disked candidates out of the original 9,200 stars. We individually inspected the remaining lightcurves. If a star was selected by these criteria but (a) had no photometric variability greater than noise (i.e. $S<1$), (b) had a $JHK$ color trajectory consistent with Gaussian noise around a mean value, and (c) on average, lay on or to the left of the boundary between region ``P'' and ``D'' in $JHK$ color space (see Fig. \ref{fig:color}), then we concluded it was not clearly a CTTS that possessed a $K$-band excess, and removed it from our analysis. 12 stars were removed this way.

\section{Results }
\label{results}

After applying these criteria we recover 30 pre-main sequence stars, whose properties we present in Tables \ref{tbl:1} and \ref{tbl:2}. We designate them RWA 1--30. 


\subsection{Identification of pre-main sequence stars}
\label{identification}

Based on the method presented in \S \ref{disks}, we report the identification of 30 young stellar objects that possess a near-infrared excess consistent with an optically thick disk at $K$-band (a ``$K$-band excess'').
Of these, 5 were previously reported as actively accreting YSOs by
\citet{Asp09}
based on Br$\gamma$ emission and other spectral signatures, and one was reported as a possible but unconfirmed YSO; the remaining 24 are new discoveries.

The positions of stars identified with $K$-band excess were checked against the IPAC database; all had a 2MASS counterpart within 0.2 arcsec, except for RWA 28 which had no counterpart within 2\arcsec. Two stars were also found as IRAS sources, and seven are AKARI sources.  Six stars have been discussed by 
\citet{Asp09}.
2MASS photometry corroborate the presence of a $K$-band excess at a significant level for 15 stars. In 10 more, the colors are within 2.5 $\sigma$ of the line separating regions ``P'' and ``D'', so would be considered ambiguous. Four stars have 2MASS colors indicating a significant \emph{lack} of $K$-band excess. The YSOs CN 3S (RWA 5) and CN 7 (RWA 13) did not show NIR excess at 2MASS epoch but were identified as $K$-band excess sources in these observations; the classification of CN 3S was inconclusive based on its spectrum at $1.4-2.5 \mu$m, but our identification of it as a variable star that possesses a $K$-band excess in these 2008-2009 observations, supports its status as a YSO. 


With the exception of source CN 3N, we recovered all of the YSOs identified in
\citet{Asp09}
that our search was sensitive to -- the only other stars that we missed were either too bright or too faint for our search, or did not show a NIR excess at the time of the 2MASS observations presented in 
\citet{Asp09}.
The recovery of spectroscopically confirmed young stars in our analysis provides a useful indication that our search is finding real YSOs. However, this is not expected to be a complete sample of all of the young stars in the field for three reasons.
First, not all stars that possess an accreting circumstellar disk (Class II stars) are identifiable in a $JHK$ color-color diagram, especially those seen at unfavorable inclination angles, low accretion rates, and/or large inner disk holes 
\citep{Mey97}.
Many of these disked stars can be recovered using longer-wavelength observations 
\citep{Hai00,Lad00}.
Second, the brightness cutoffs used in this study to guarantee reliable photometry exclude the brighter PMS stars and, if they exist, fainter or more substantially extincted ($A_V > 7$) low-mass stars. 
Three stars in this field (CN 3N, Cyg 19, IRAS 15N) have been confirmed as YSOs based on spectroscopic and 2MASS observations 
\citep{Asp09},
but are saturated in the WFCAM images; 
two confirmed PMS stars (the Braid Star and IRAS 14) are likewise fainter than our cutoff.
Finally, in this analysis we removed all stars that showed any photometric error flags, such as from deblending or bad pixels, that may in fact have useful photometry sufficient to identify a $K$-band excess. Nonetheless, our goal in this paper was not a complete determination of the YSO population, but rather a high-confidence sample of $K$-excess stars whose NIR variability properties could be reliably studied.

\subsection{Variability of pre-main sequence stars}
\label{pms-var}
\begin{figure}
\plotone{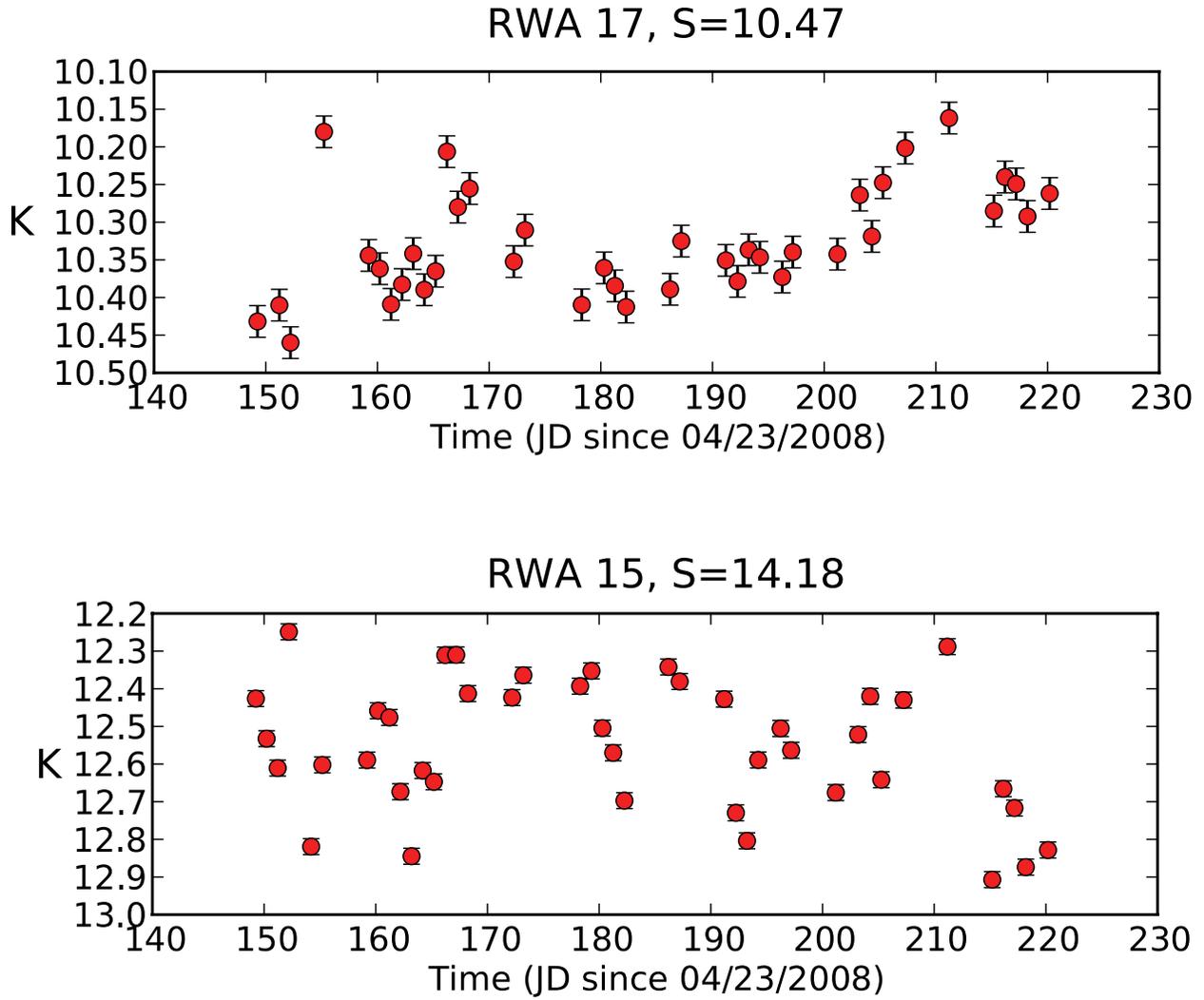}
\caption{K band light curves for 2 sample stars RWA15 and RWA 17.  Data are from season 2. The $Stetson$ index for each star is given for season 2.}
\label{new}
\end{figure}

Of the 30 YSOs, 28 (93\%) are variable at a significant level.
Values of $S$ among these stars range from $S=2$ to $S=60$. Variable stars typically vary in all 3 bands, with most stars also showing color variations. $J$-band RMS variability in these stars ranged widely from 0.02 mag to 0.70 mag, peak-to-trough variability ranging from near the photometric noise limit to greater than two magnitudes at $J$ (Table~3). The median $J$-band RMS on these variable YSOs was $\sim$0.1 mag, corresponding to a median variability index $S\sim10$.  YSOs  varied significantly on all timescales studied. Many varied noticeably from one night to the next.  However, the manner of the variations differed among the stars with some showing slow and steady changes, while others were more abrupt.   Figure~\ref{new} shows the $K$-band lightcurves from season 2 for two ``typical'' stars, RWA 15 and RWA 17.  In the case of RWA 15, the global range is about 0.75 mag with night to night changes of nearly 30\%.  The data also seem to have a pattern of peaks and troughs separated by about 10 days (these will be discussed further in paper~II).    RWA 17 is a little chaotic in the beginning, but in general, it shows a slow steady increase of 25\% over the course of a month, followed by a decline.

The only two $K$-excess sources which are not identified as variable, RWA 28 and 30, are both at the faintest end of our search near $J=17$ with the largest photometric uncertainties, typically around 4\% at $J=17$; hence the 2\% variability noted in some brighter stars could go undetected here. RWA 30 is plausibly variable under its photometric noise: its $S$ value is 0.76,
$3.5 \sigma$ higher than the mean $S=0.2$ value for non-variable stars,
and its observed $J$-band RMS value (while dominated by the photometric uncertainty) is higher than four disked stars identified as variable. 
RWA 28, on the other hand, shows no indications of any true variability: its variability index $S=0.18$ is consistent with stars whose observed variations arise purely from photometric noise. 
RWA 28's photometric noise causes it to drift near the border between ``P'' and ``D'' (in similar fashion to the 12 stars rejected from our source list as described in \S \ref{transient}). 
It was not excluded from our source list because unlike the 12 excluded stars, RWA 28's mean $JHK$ colors lie squarely in region ``D''. 
No other stars in our source list are suspicious in this way. 
Overall, over 90\%  of the YSOs we identify are variable at a significant level.


\subsection{Extreme  NIR excess.}
\label{classification}
Twenty-five stars lie in region ``D'' for the color-color diagram for at least part of the observations, 
The remaining 5 stars, which lie in ``E'', have more excess at $K$-band than can be accounted for by accreting T-Tauri stars, like those in Taurus which were used by \citet{Mey97} to derive the cTTSs locus.  We refer to these as ``extreme $K$-excess stars''.
Four of the five extreme $K$-excess stars (RWA 2, 15, 19, and 26) exhibit extreme variability as well, with variability index $S>15$. 
These stars may be younger, more active counterparts to the relatively quiet classical T Tauri stars that inhabit region ``D''; the extra $K$-band excess may arise from warm, infalling circumstellar material that is not in a disk. 
Indeed, three of these stars are detected as AKARI ($9-200 \mu$m) sources, and the brighter two are also IRAS ($25-100 \mu$m) sources.  Spectral energy distribution fits of these mid and far IR data following \citet{Rob06} support the interpretation that they are less-evolved ``Class I'' protostars. Two of these stars also give indications of being eruptive variables (Wolk et al. 2012, in prep).

\subsection{Transient NIR excesses.}
\label{transient-results}

\begin{figure}
  \plotone{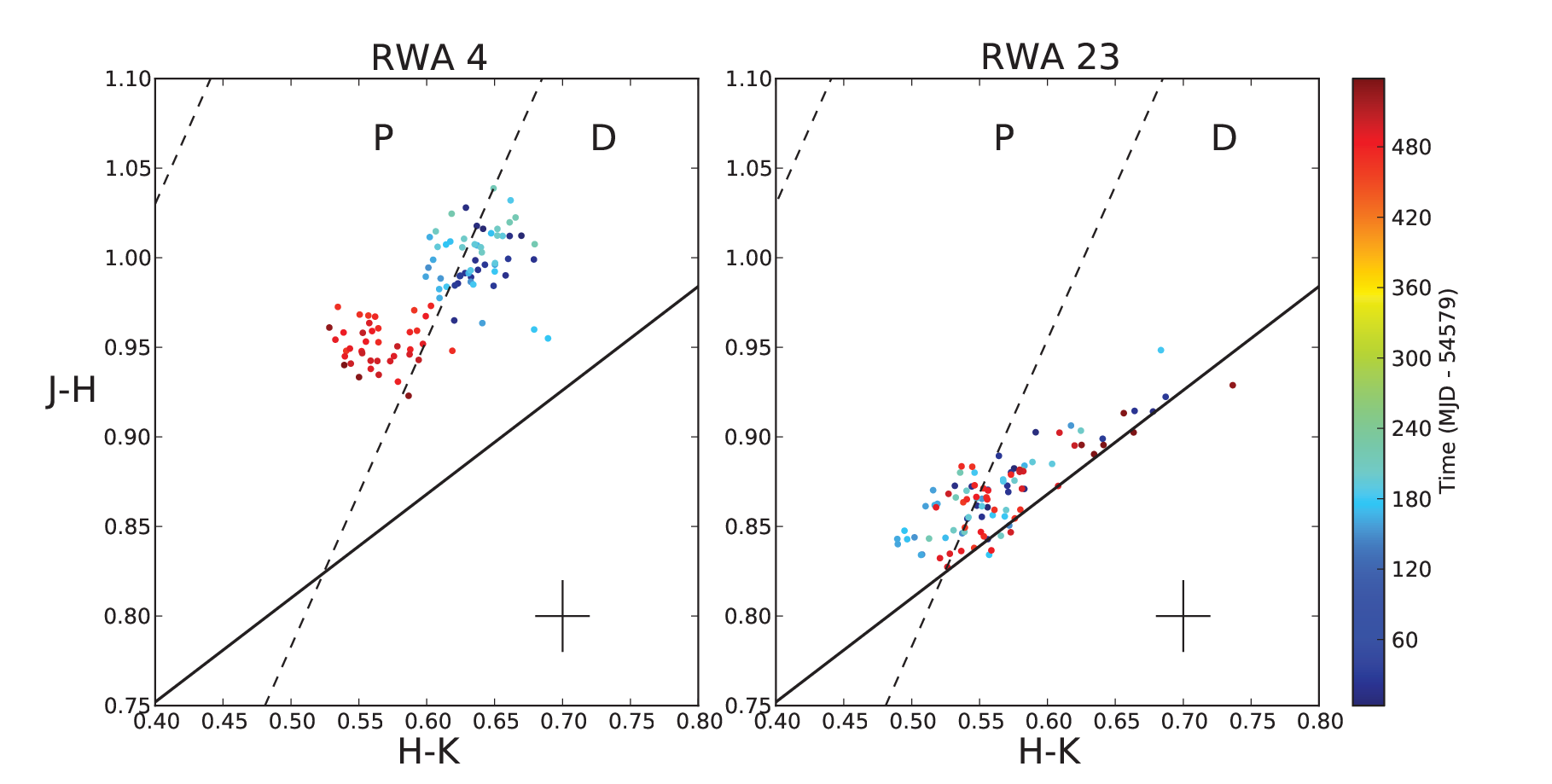}
  \caption{ 
    $JHK$ color trajectories for RWA 4 and 23, two of the nine YSOs 
    identified in this analysis as having a transient near-infrared excess.
    RWA 4 has a significant near-infrared excess in only 41\% of observations,
    and its time-averaged mean $JHK$ colors lie in region ``P''. RWA 23
    exhibits a significant NIR excess in 59\% of observations. Colored circles
    indicate the progression of time from early 2008 (dark blue) to late 2009
    (dark red).
    Solid line: CTTS locus. Dashed line: reddening vector. The plus (+) in the
    bottom-right corner illustrates the typical uncertainty on each individual
    $JHK$ measurement.
    \label{fig:track2}
  }
\end{figure}

Of the 25 simple $K$-excess stars, seven vary in color space such that they spend more than 15\% of their time in region ``P'', and would not be detected by near-infrared excess criterion at these epochs (see Fig. \ref{fig:hist}).  Figure~\ref{fig:track2} shows examples of two such stars.
Further, 3 of these 7 stars have mean colors that lay in region ``P''; these YSOs would be undetected in a search of time-averaged $JHK$ color.
Finally, comparison with 2MASS data show two stars that possess a $K$-band excess in all of these UKIRT observations but show no significant $K_s$-band excess at the 2MASS epoch. 
These nine stars (nearly 1/3 of our sample), have been identified as exhibiting a transient $K$-band excess.

\begin{figure}
  \plotone{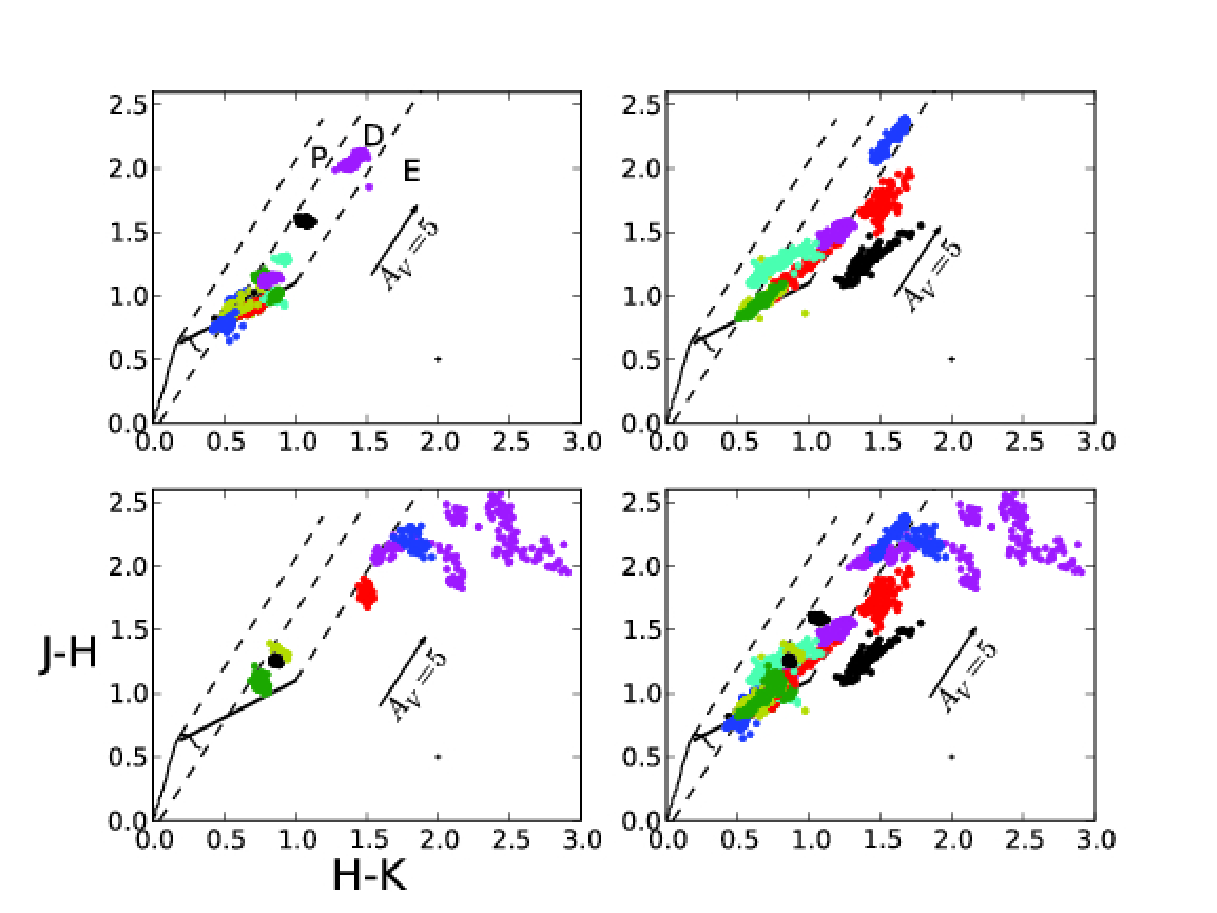}
  \caption{
  The color trajectories of 30 YSOs over the course of the observations. 
 Each starÕs trajectory is plotted in a different color; some colors are repeated. 
Color trajectories can be broadly divided into three groups:  small systematic variations  (upper-left), large systematic variation (upper-right), and others -- including large stochastic variables and a few stars that parallel the cool main sequence (lower-left). All are overlaid in the lower-right.
Nine YSOs drift between regions ÒPÓ (photosphere) and ÒDÓ (disk) and could be missed by single-epoch observations.
A few lie, on average, in region ÒPÓ and would likely be invisible to a search of time-averaged
JHK color. The small plus (+) in the bottom-right of each panel illustrates the typical uncertainty on each individual measurement.
    \label{fig:track30}
  }
\end{figure}

Among the variable CTTS candidates, many show $JHK$ color-color variability parallel to either the CTTS locus 
\citep{Mey97}
or to a combination of the CTTS locus plus extinction; the two remaining show chaotic behavior in $JHK$ space. 
In Figure~\ref{fig:track30} we show the trajectories of the thirty YSOs. The upper-left panel shows the simple disked variables which show small ($<$ 0.5 mag in color) variations which, for the most part, appear to move the star parallel to the main sequence track or directly along the CTTS locus.  
The lower-left panel shows the extreme variables, plus a few stars which move parallel to the main sequence.  The upper-right panel shows eight trajectories which appear to be indicative of systematic changes in the disk structure.  
Theoretical models of the CTTS locus 
\citep{Mey97}
derive its slope as owing to different accretion rates, disk hole sizes, and inclination angles.
Among the extreme $K$-excess stars, color-space variability is largely chaotic, but in two cases seems to roughly follow the same pattern of positive color slope that seems to contain contributions from the CTTS track and from the dust reddening track.  Of course, there are more than just 3 parameters (accretion rates, disk hole size, and inclination angle) which determine the final location.  By varying 14 parameters  in their radiative transfer--based models, \citet{Rob06}  calculate 200,000 model SEDs in evolutionary stages.\footnote{ \citet{Rob06} use ``stages'' as a theoretical equivalent to the  observational ``classes''  but the mapping is not exact since stages 0-III cover Classes 0-II. The Class of an object can depend both on Stage and, for example, viewing angle.}   Additional model parameters that appear susceptible to short timescale variations include the effective stellar temperature, which can change due to flares or spots,  as well as parameters relating to the disk structure such as the scale height of the inner disk. 

While we see stars regularly cross between ``P'' and ``D'', no stars cross between ``D'' and ``E''.  Our sample is very small and not cleanly defined in terms of Class.   thus, the results are more open to speculation than interpretation.  The YSOs  in this sample seem to separate into simple-disked Class II stars that inhabit regions ``P'' and ``D'', and more extreme sources that inhabit region ``E''.   Models describe all but one of the stars in the ``E'' region as stage I  \citep{Rob06}. There is a paucity of stage I models which occupy the ``D'' region.   This supports speculation that these are Class I sources and we infer from \citet{Rob06} that changes in the various accretion parameters in these stars lead to changes in $J-H$ and $H-K$ color which are  mediated by an envelope which is more complex than the thin disk surrounding Class II (Stage III) objects.  

\section{ Discussion }
\label{discussion}

\subsection{YSOs are variable in the near-infrared }


As found in \S \ref{pms-var}, virtually all of the detected YSOs showing a $K$-band excess also exhibit near-infrared variability. Importantly, our search did not include variability as a selection criterion except to disambiguate close cases.  
As noted in \S 3.2, our study is not complete. For example, only 6 of the 12 YSOs discussed in \citet{Asp09} were recovered by our study. Further, the stars in our sample were subject to both brightness and faintness cuts to ensure sensitivity to photometric variations on the order of $\sim$2\%. Nonetheless, it is clear
that near-infrared variability is a behavior common to all disked pre-main sequence stars bright enough to be measured at $>$2\% accuracy and possess a $K$-band excess. This is consistent with previous NIR variability studies of young stars 
\citep{Car01,Car02}
and also consistent with optical studies 
\citep{Her94,Bri05}.

It seems likely that NIR variability could be used alone to identify young stars, as seen in the optical
\citep[e.g.][]{Bri05, Par09}. \citet{Par09} noted that long term monitoring increased the variability detection rate in the optical by about 50\% for periodic variables. 
We do not find the effect of extended monitoring as pronounced. Twenty-six of the 30 stars were found to be variable via the Stetson index in the 26 night, first observing season.  For the 70+ nights of seasons two and three, the results were 25/30 and 27/30 respectively.  Even the inclusion of all three seasons only lead to the detection of 28/30 as variables.  Because it was consistently the same stars which were detected as non-variable (RWA 6, 8, 11, 28 and 30),  it appears the detection of variability on the longer datasets was not  an effect of long term periodicity, but rather the increase in signal to noise enabled by the additional data.
We suspect that  using the specific trajectories in $JHK$ color space seen in these 30 stars (\S \ref{transient-results}) 
as an additional selection criterion could be useful in detecting disked stars within the ``P'' portion of the reddening band.
This would also be consistent with variability seen in 
\citet{Car01}
where stars that lack near-infrared excess, but that are associated with the Orion A molecular cloud, are seen to vary at a level significantly higher than field stars. This means that Class III stars should be detected as NIR variables \citep[e.g.][]{Par09, Wol12}.

\subsection{On the variability of the NIR disk diagnostic }

While other studies have used time-series $JHK$ photometry to investigate young stars with disks, the use of time-averaged NIR colors to identify disked stars 
\citep[e.g.][]{Car02}
 will still miss some YSOs. 
In this study, we found three stars whose time-averaged colors showed no infrared excess, but whose variability carried them into region ``D'' of $JHK$ space in 20\%-50\% of observations, revealing the presence of a circumstellar disk around these stars. Therefore, simply searching through time-averaged colors is not a sufficient YSO detection technique in time-series observations.


$JHK$ observations are known not to be sensitive to 100\% of disks around young stars. The CTTS locus is partially degenerate with reddened main sequence colors in $JHK$ space 
\citep{Mey97},
and previous infrared studies of young stellar populations show $L$-band (3.5 $\mu$m) observations can detect disks around $\sim 85 \%$ of young stars at age $\sim 0.3 - 1$ Myr,  while
$JHK$-only single-epoch surveys see disks around only $\sim 50-60 \%$ of the same sample of stars
\citep{Hai00,Lad00}.
We summed up the probability of seeing a disk around each of the RWA stars on a given night.
The probabilities range from $\sim$ 18\% through 80\% with many stars that always showed disks  (100\%).
We then calculated an expectation value of how many disked stars we expect to see on a single night. 
For our data this came out to about 25. So on an average night we expect to see 25 of the 30 RWA stars in the ``D" region of the diagram. 
Multiple observations gave us 30 stars, i.e.\ an $\sim$20\% increase.
If our results are typical, then a direct consequence of this study is that 20$\pm$ 8\% more disked stars may be found by using multiple $JHK$ observations spread out over about a month, increasing  $JHK$ 
disk sensitivity to roughly $60\% - 70\%$. In situations where it is significantly more practical to obtain multiple $JHK$ observations than to acquire $L$-band imaging or to carry out a spectroscopic survey to investigate accretion, this approach could prove a useful way to simultaneously increase the number of identified circumstellar disks and study variability of young stars.

\subsection{On the underlying cause for NIR variability in YSOs }

Of the 30 YSOs, 28 are variable at a significant level. As seen in the upper portion of Figure~\ref{fig:track30}, about half of these vary along linear tracks. Some YSO's parallel the CTTS locus of 
\citet{Mey97}, others seem follow a somewhat steeper slope. 
As presented in \S \ref{transient-results}, 
the aggregate color-domain variability behavior is consistent with changes in mass accretion rate, inner hole size, and inclination angle, in some cases combined with changes in extinction or starspot coverage. 
Other variability mechanisms exist.  These were summarized recently by \citet{Sch09}.  The dominant process can be indicated by the range of magnitude and color changes exhibited by the stars (summarized in Table~\ref{tbl:3}). Among these mechanisms are rotationally modulated changes due to cool spots or hot spots on the stellar surface, extinction changes, and changes in the inner disk.  Our goal in this section is to discuss possible factors that may induce the observed variability, not to distinguish among them.

\begin{figure} 
\includegraphics[]{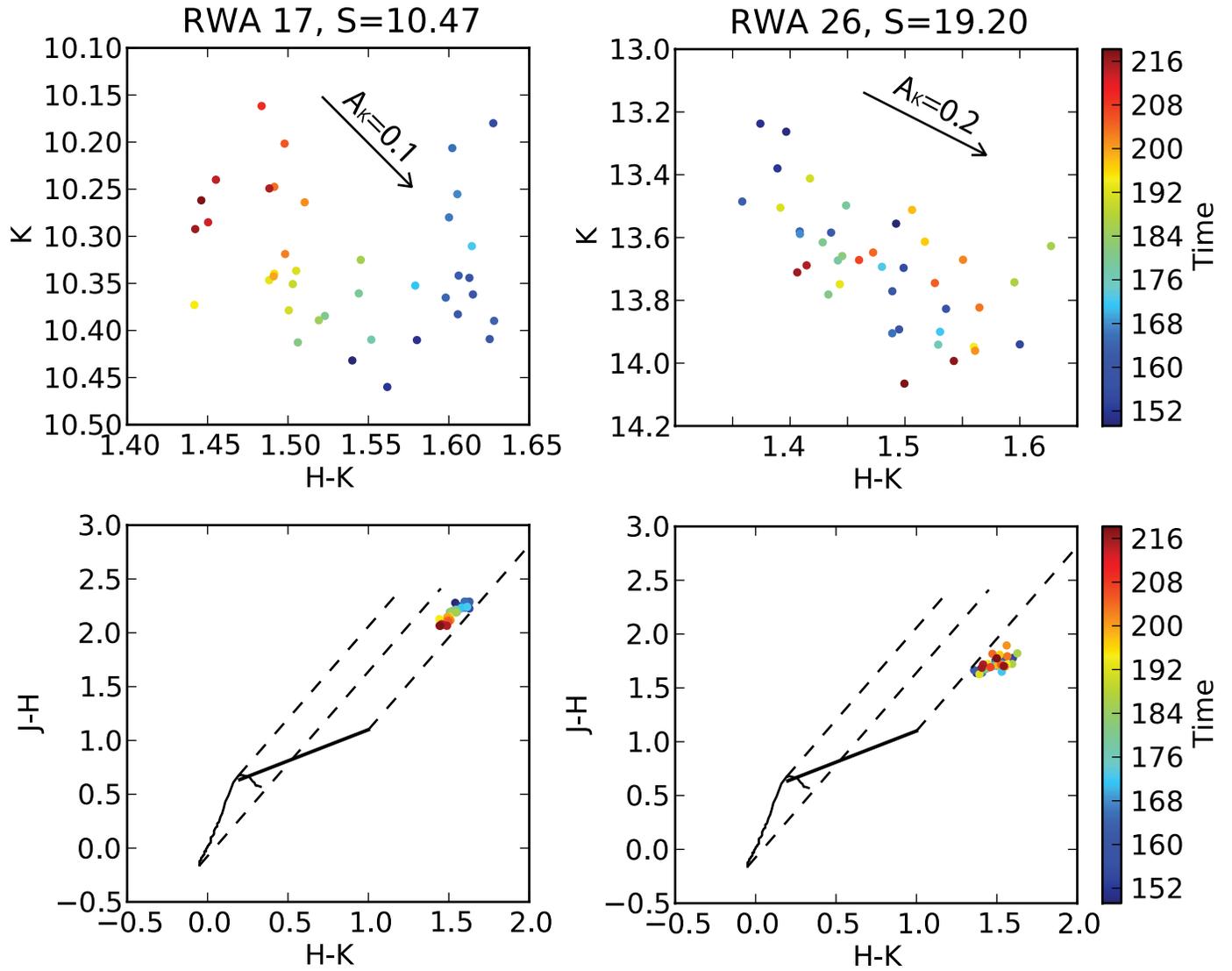}
\caption{Season 2 color data for RWA~17 (left) and RWA ~26 (right).  While the data generally track the reddening vectors, the significant width of the tracks indicates a secondary cause of the variations. ``Time'' refers to days since the first observation -- April 26, 2008}
\label{red} 
\end{figure}

  Changes in the overall extinction may be the simplest to imagine. Perhaps induced by the disk, extinction can cause unlimited changes in the apparent flux of the stars.  However, such changes should move the star in the direction of the reddening vector. Figure~\ref{fig:track30} shows no pure examples of this. However, there are many cases where the data appear to move predominantly in this direction (see Figure~\ref{fig:track30} {\it upper-right}).  RWA~17 and RWA~26 show some of the clearest examples of changes in reddening  (Fig. \ref{red}).  However, it is clear from the color--magnitude plots that the observed changes are not due to reddening alone.  

Cool spots, like those on the Sun, were first identified as a contributor to the variability of PMS stars in the 1980s \citep{Vrb85}.
Even static stellar spots induce variability because of the rotation of the star. Starspots have been used  regularly as a method of measuring stellar periods \citep[e.g.][]{Att92}.  But there is a limit to the variability cool spots can induce, since the spot is typically only  1000-1500K cooler than the nominal photosphere.  In the I band, the luminosity change is typically $<$ 15\% \citep{Coh04}.  
The implied color change due to a lower effective temperature is $<$ 5\%.  
All the stars in our sample exceed a color range of 9\% in $J-K$ (Table~\ref{tbl:3}).

Hot spots, thought to arise from accretion, can cause a larger signal than cool spots since the temperature difference is typically larger (a factor of 2 or 3 hotter than the surrounding photosphere).  These can induce signals as high at 1 magnitude at $J$ and color changes of 40\% in $J-K$, even with a filling factor as small as 1\% \citep{Sch09}. However, over 1/3 of our sample exceeds this color range, so hot spots alone cannot account for this variability.  Of the remaining 20 stars,  half of them have color changes in excess of of 25\% in $J-K$, indicative of very active hot spots or a combination of variable hot spot and other effects.

The CTTS locus is derived from models of T Tauri stars with accretion rates spanning two orders of magnitude, 
disk hole sizes spanning 
$1-10 R_\star$, 
and a full range of observable inclination angles 
\citep[see][esp. Fig. 3 and Fig. 18 respectively]{Mey97, Rob06}.
That most of the $JHK$ variability we see in CTTS candidates is focused along this track is evidence that changes in the overall accretion structure --
disk inclination, hole size and accretion rate (the size of the hot spots) -- are the primary cause for $JHK$ variability in  about half  of the stars.  This is especially true for the subset of stars in  Figure~\ref{fig:track30} {\it upper-left} which move right along this track and those in Figure~\ref{fig:track30} {\it upper-right} which appear to follow a hybrid of this track plus reddening.

Because of the degeneracy between the 3 parameters (disk inclination, hole size and accretion rate), it is not possible to easily disentangle the contributions from each of these 3 parameters. That said, it is easy to imagine ways that they might co-vary. For example, a decreasing (or increasing) inner disk hole size might naturally be simultaneous with an increasing (or decreasing) accretion rate.
The line-of-sight inclination of the innermost edge of the disk -- not the inclination of the entire disk -- might reasonably vary due to warping in response to a strong, misaligned stellar magnetic field and the rotation of the star. Observational and theoretical evidence for warped accretion disks has been provided by 
\citet{Bou03} and \citet{Esp11}.
 For the data presented here, we do not attempt to model the individual stars to identify the specific mechanisms of variability. 
 

\subsection{Individual variability }

In addition to the aggregate color-domain variability just analyzed, we have identified a number of striking pattens of variability in individual stars' lightcurves. Periodic, quasi-periodic and eruptive variability is seen among the identified YSOs, mirroring previously studied classes of variable YSOs such as the periodic disk eclipses of AA Tau 
\citep{Bou03},
and the eruptive, large-scale accretion events of EX Lup and V1118 Ori
\citep{Asp10,Aud10}.  
Many classes of variability including eclipsing and contact binaries, and ``long-period'' ($P\sim$ weeks) pulsating stars are seen among the $\sim 160$ variable field stars.  In one case one of the disked stars appears to be part of an eclipsing system.  A detailed investigation of these variable stars will appear in Paper II.

\section{Summary}
\label{summary}

We observed a star-forming region in the dark cloud L 1003 in Cyg OB7 on more than 100 nights spanning 1.5 years using NIR wide-band photometry. Using the $K$-band excess diagnostic, we found 30 candidate PMS stars, including 25 disked objects (CTTS candidates) and 5 young stars with extreme $K$-band excess (Class~I candidates).
Among the 25 CTTS candidates, nine (36\%) cross the main sequence reddening band cutoff, indicating that single-epoch observations are insufficient to identify all YSOs that show $K$-band excess. Even time-series observations may miss some stars if they only select using time-averaged $JHK$ colors. Additionally, the pattern of variability in color space seen in the variable CTTS candidates is a strong indication that NIR variability in young stars arises from a combination of variable extinction and changes in the inner accretion disk. 
While some of the variability may be due to rotationally modulated starspots other possibilities include changes in accretion rate, inner hole size, and/or disk inclination. None of the extreme $K$-band excess stars are seen to cross into the ``disked'' region of the NIR color-color space.

To summarize our results: \\
(1) The 30 pre-main sequence stars discussed in this paper include 24 newly identified YSOs. \\
(2) Overall, $>$90\% of the young stars with disks are variable. Over 80\% are variable on a time scale of about 1 month.  \\
(3) YSOs can be separated into ``simple-disk'' or ``extreme'' classes based on degree of $K$-band excess. \\
(4) 36\% of ``simple-disk'' $K$-band excess sources have a transient $K$-band excess. \\
(5) The color behavior of many of the ``simple-disk'' YSOs is consistent with changes in disk geometry and/or accretion rate. \\

In this paper we have presented an analysis of a unique dataset: containing multi--season NIR monitoring for variability of young stars. A follow-up paper \citep{Wol12} will discuss the variability of field stars, and a phenomenological categorization of NIR variability seen in YSOs. Further observations, at both IR and X-ray wavelengths, are planned to better characterize the overall pre-main sequence population in this field.

\section{Acknowledgements}


Thanks to Joseph Hora and David Charbonneau for useful comments as this research project was being developed. Thanks also to Mike Read and Nicholas Cross for assistance with the data retrieval. Thanks to Bo Reipurth for stimulating discussions.

S.J.W. is supported by NASA contract NAS8-03060 (Chandra). T.S.R.  was supported by Grant \#1348190 from the Spitzer Science Center.
Thanks also to the NSF REU program for funding part of this research via NSF REU site grant \#0757887.
This research has made use of the NASA/ IPAC Infrared Science Archive, which is operated by the Jet Propulsion Laboratory, California Institute of Technology, under contract with the
National Aeronautics and Space Administration.
The United Kingdom Infrared Telescope is operated by the Joint Astronomy Centre on behalf of the Science and Technology Facilities Council of the U.K. We thank A. Nord, L. Rizzi, and T. Carroll for assistance in obtaining these observations. We also thank the University of Hawaii Time Allocation Committee for allocating the nights during which these observations were made. The authors wish to recognize and acknowledge the very significant cultural role and reverence that the summit of Mauna Kea has always had within the indigenous Hawaiian community. We are most fortunate to have the opportunity to conduct observations from this sacred mountain.







\tighten


\begin{deluxetable}{llllllll}

  \tabletypesize{\scriptsize}
  
  \tablecaption{Object Coordinates, Identification, and Median Photometry\label{tbl:1}}
  \tablewidth{0pt}

  \tablehead{
    \colhead{ } &
    \colhead{ } &
    \colhead{ } &
    \colhead{ } &
    \multicolumn{3}{c}{Median WFCAM Photometric Values} &
    \colhead{ } \\
    \cline{5-7} \\
    
    \colhead{Object ID} &
    \colhead{R.A. (J2000)} & 
    \colhead{Decl. (J2000)} & 
    \colhead{2MASS ID} &
    \colhead{$J$ } &
    \colhead{$H$ } &
    \colhead{$K$ } &
    \colhead{Other ID} 
  }
   \startdata
  RWA 1 & 21:00:16.570 & +52:26:23.105 & 21001656+5226230 & 12.886$\pm$0.021 & 11.796$\pm$0.021 & 10.982$\pm$0.021 & CN 2\tablenotemark{a} \\
  RWA 2 & 21:02:05.456 & +52:28:54.477 & 21020547+5228544 & 16.487$\pm$0.028 & 14.155$\pm$0.022 & 11.851$\pm$0.021  &21005+5217\tablenotemark{b} \\
 ~ & ~ & ~& ~ & ~ & ~ &  ~  &  2102055+522854\tablenotemark{c} \\
  RWA 3 & 20:59:04.194 & +52:21:44.936 & 20590419+5221448 & 14.404$\pm$0.022 & 13.098$\pm$0.021 & 12.208$\pm$0.021 \\
  RWA 4 & 20:58:59.790 & +52:22:18.340 & 20585978+5222182 & 12.403$\pm$0.021 & 11.420$\pm$0.021 & 10.823$\pm$0.021 \\
  RWA 5 & 21:00:05.086 & +52:34:04.916 & 21000508+5234049 & 12.875$\pm$0.021 & 11.601$\pm$0.021 & 10.778$\pm$0.021  &CN 3S\tablenotemark{a} \\
  RWA 6 & 21:02:43.791 & +52:23:48.278 & 21024378+5223484 & 14.865$\pm$0.022 & 13.605$\pm$0.021 & 12.741$\pm$0.021 \\
  RWA 7 & 21:00:02.165 & +52:35:16.205 & 21000217+5235160 & 15.361$\pm$0.023 & 13.872$\pm$0.021 & 12.646$\pm$0.021  &2100019+523515\tablenotemark{c}\\
  RWA 8 & 21:02:14.992 & +52:32:40.380 & 21021498+5232405 & 16.114$\pm$0.025 & 15.107$\pm$0.023 & 14.400$\pm$0.023 \\
  RWA 9 & 21:02:38.301 & +52:17:40.841 & 21023831+5217408 & 14.758$\pm$0.022 & 13.620$\pm$0.021 & 12.872$\pm$0.021 \\
  RWA 10 & 21:00:55.768 & +52:25:22.066 & 21005576+5225221 & 14.783$\pm$0.022 & 13.833$\pm$0.021 & 13.188$\pm$0.021 \\
  RWA 11 & 20:59:46.192 & +52:33:21.470 & 20594619+5233216 & 13.159$\pm$0.021 & 11.881$\pm$0.021 & 10.961$\pm$0.021 \\
  RWA 12 & 20:59:51.844 & +52:40:20.613 & 20595184+5240205 & 14.190$\pm$0.021 & 12.132$\pm$0.021 & 10.739$\pm$0.021 &  2059518+524020\tablenotemark{c} \\
  RWA 13 & 21:00:17.257 & +52:28:25.488 & 21001725+5228253 & 12.100$\pm$0.021 & 11.199$\pm$0.021 & 10.632$\pm$0.021 & CN 7\tablenotemark{a} \\
  RWA 14 & 21:02:01.476 & +52:54:35.758 & 21020145+5254357 & 15.476$\pm$0.023 & 14.632$\pm$0.022 & 14.087$\pm$0.022 \\
  RWA 15 & 21:00:35.175 & +52:33:24.410 & 21003517+5233244 & 15.128$\pm$0.022 & 13.883$\pm$0.021 & 12.432$\pm$0.021 & CN 1\tablenotemark{a},  20590+5221\tablenotemark{b}  \\
   ~ & ~ & ~& ~ & ~ & ~ &  ~  &2100352+523324\tablenotemark{c} \\
  RWA 16 & 20:58:49.664 & +52:19:46.513 & 20584965+5219465 & 12.749$\pm$0.021 & 11.862$\pm$0.021 & 11.249$\pm$0.021 \\
  RWA 17 & 20:58:50.100 & +52:32:54.427 & 20585009+5232543 & 14.221$\pm$0.021 & 11.957$\pm$0.021 & 10.383$\pm$0.021 &  2058500+523255 \\
  RWA 18 & 20:59:45.016 & +52:17:49.147 & 20594500+5217492 & 12.756$\pm$0.021 & 11.832$\pm$0.021 & 11.115$\pm$0.021 \\
  RWA 19 & 20:59:40.710 & +52:34:13.470 & 20594071+5234135 & 14.679$\pm$0.022 & 12.460$\pm$0.021 & 10.655$\pm$0.021 & CN 6\tablenotemark{a}\\     
  ~ & ~ & ~& ~ & ~ & ~ &  ~  &2059408+523414\tablenotemark{c} \\
  RWA 20 & 21:00:19.042 & +52:27:28.307 & 21001903+5227281 & 11.622$\pm$0.021 & 10.631$\pm$0.021 & 9.794$\pm$0.021 & CN 8\tablenotemark{a}\\ 
  ~ & ~ & ~& ~ & ~ & ~ &  ~  &2100191+522728\tablenotemark{c} \\
  RWA 21 & 21:00:36.443 & +52:12:03.049 & 21003643+5212030 & 12.748$\pm$0.021 & 11.807$\pm$0.021 & 11.164$\pm$0.021 \\
  RWA 22 & 20:59:08.952 & +52:22:39.320 & 20590895+5222392 & 13.197$\pm$0.021 & 12.066$\pm$0.021 & 11.244$\pm$0.021 \\
  RWA 23 & 20:57:43.462 & +52:51:00.401 & 20574345+5251004 & 14.892$\pm$0.022 & 14.040$\pm$0.021 & 13.476$\pm$0.021 \\
  RWA 24 & 20:59:19.282 & +52:25:43.857 & 20591928+5225437 & 13.657$\pm$0.021 & 12.077$\pm$0.021 & 11.008$\pm$0.021 \\
  RWA 25 & 21:01:02.586 & +52:27:08.640 & 21010258+5227086 & 11.859$\pm$0.021 & 11.085$\pm$0.021 & 10.610$\pm$0.021 \\
  RWA 26 & 21:01:02.575 & +52:24:00.053 & 21010256+5223599 & 16.913$\pm$0.036 & 15.199$\pm$0.023 & 13.707$\pm$0.022 \\
  RWA 27 & 21:01:16.921 & +52:28:32.587 & 21011691+5228325 & 14.704$\pm$0.022 & 13.514$\pm$0.021 & 12.832$\pm$0.021 \\
  RWA 28 & 20:59:51.642 & +52:41:32.479 & not detected & 16.922$\pm$0.037 & 15.837$\pm$0.028 & 15.080$\pm$0.026 \\
  RWA 29 & 20:59:10.396 & +52:07:44.327 & 20591038+5207442 & 16.115$\pm$0.025 & 15.124$\pm$0.023 & 14.258$\pm$0.022 \\
  RWA 30 & 21:01:01.073 & +52:10:42.787 & 21010106+5210427 & 17.049$\pm$0.039 & 15.275$\pm$0.024 & 13.775$\pm$0.022 \\

  \enddata
   \tablecomments{ Median photometric values were extracted from 100 $JHK$ observations of each star. }
  \tablenotetext{a}{From Aspin et al.\ 2009.}
  \tablenotetext{b}{AKARI ID}
  \tablenotetext{c}{IRAS ID}
 
\end{deluxetable}




\begin{deluxetable}{l@{\extracolsep{10pt}}
    r@{\extracolsep{10pt}}r@{\extracolsep{10pt}}r@{\extracolsep{10pt}}
    c@{\extracolsep{0pt}}c@{\extracolsep{10pt}}
    c@{\extracolsep{0pt}}c@{\extracolsep{0pt}}c}

  \tabletypesize{\scriptsize}
  
  \tablecaption{Variability Characteristics\label{tbl:2}}
  \tablewidth{0pt}

  \tablehead{

    \colhead{ } &
    \multicolumn{3}{c}{Observed RMS}    & 
    \multicolumn{2}{c}{Color RMS}  & 
    \colhead{Stetson  index} & 
    \colhead{P/D/E} & 
    \colhead{Transient excess?} \\
    \cline{2-4} 
    \cline{5-6} \\
    \colhead{Object ID} &
    \colhead{$J$ } &
    \colhead{$H$ } &
    \colhead{$K$ } &
    \colhead{$J-H$} &
    \colhead{$H-K$} &
    \colhead{$S$} &
    \colhead{(on average)} 
  }
  
  \startdata

  RWA 1 & 0.473 & 0.379 & 0.278 & 0.099 & 0.103 & 42.96 & D & no \\
  RWA 2 & 0.698 & 0.746 & 0.493 & 0.191 & 0.361 & 59.95 & E & no \\
  RWA 3 & 0.022 & 0.022 & 0.019 & 0.021 & 0.019 & 3.78 & D & no \\
  RWA 4 & 0.086 & 0.085 & 0.087 & 0.028 & 0.041 & 8.59 & P & yes \\
  RWA 5 & 0.064 & 0.051 & 0.054 & 0.023 & 0.028 & 7.72 & D & yes\tablenotemark{a} \\
  RWA 6 & 0.020 & 0.023 & 0.031 & 0.013 & 0.013 & 2.23 & D & yes\tablenotemark{a} \\
  RWA 7 & 0.241 & 0.276 & 0.295 & 0.059 & 0.068 & 30.81 & D & no \\
  RWA 8 & 0.101 & 0.056 & 0.067 & 0.050 & 0.068 & 2.77 & D & yes \\
  RWA 9 & 0.066 & 0.078 & 0.094 & 0.017 & 0.021 & 8.65 & D & no \\
  RWA 10 & 0.028 & 0.032 & 0.043 & 0.016 & 0.021 & 3.33 & D & no \\
  RWA 11 & 0.029 & 0.027 & 0.041 & 0.014 & 0.016 & 3.40 & D & no \\
  RWA 12 & 0.154 & 0.133 & 0.134 & 0.037 & 0.042 & 14.45 & D & no \\
  RWA 13 & 0.093 & 0.087 & 0.091 & 0.030 & 0.042 & 7.95 & D & yes \\
  RWA 14 & 0.265 & 0.201 & 0.184 & 0.050 & 0.072 & 25.05 & P & yes \\
  RWA 15 & 0.248 & 0.199 & 0.184 & 0.104 & 0.110 & 16.58 & E & no \\
  RWA 16 & 0.127 & 0.112 & 0.091 & 0.028 & 0.048 & 11.50 & D & no \\
  RWA 17 & 0.287 & 0.194 & 0.137 & 0.098 & 0.070 & 24.21 & D & no \\
  RWA 18 & 0.135 & 0.115 & 0.112 & 0.037 & 0.053 & 11.53 & D & no \\
  RWA 19 & 0.135 & 0.182 & 0.145 & 0.067 & 0.057 & 16.15 & E & no \\
  RWA 20 & 0.128 & 0.126 & 0.144 & 0.030 & 0.032 & 10.80 & D & no \\
  RWA 21 & 0.245 & 0.190 & 0.136 & 0.062 & 0.068 & 19.91 & D & no \\
  RWA 22 & 0.062 & 0.067 & 0.093 & 0.019 & 0.034 & 7.59 & D & no \\
  RWA 23 & 0.054 & 0.056 & 0.073 & 0.022 & 0.043 & 6.74 & D & yes \\
  RWA 24 & 0.052 & 0.048 & 0.061 & 0.014 & 0.023 & 5.51 & D & no \\
  RWA 25 & 0.075 & 0.055 & 0.067 & 0.032 & 0.042 & 6.50 & D & yes \\
  RWA 26 & 0.270 & 0.244 & 0.205 & 0.097 & 0.078 & 15.87 & E & no \\
  RWA 27 & 0.247 & 0.181 & 0.132 & 0.070 & 0.058 & 34.58 & P & yes \\
  RWA 28 & 0.035 & 0.019 & 0.015 & 0.036 & 0.025 & 0.18 & D & no \\
  RWA 29 & 0.037 & 0.042 & 0.052 & 0.022 & 0.020 & 3.61 & D & no \\
  RWA 30 & 0.033 & 0.020 & 0.022 & 0.032 & 0.019 & 0.76 & E & no \\
 
  \enddata

  \tablenotetext{a}{Transient $K$-excess classification based on 2MASS data}
  \tablecomments{ Typical photometric errors are $\sim2\%$ Refer to Table 1 for more details. }

\end{deluxetable}




\begin{deluxetable}{lcccccc} 
  
  \tablecaption{Variability extrema \label{tbl:3}}
  \tablewidth{0pt}

  \tablehead{

    \colhead{Object ID} &
    \colhead{Median $K$ } &
    \colhead{$\Delta J$ } &
    \colhead{$\Delta K$ } &
    \colhead{$\Delta J-H$} &
    \colhead{$\Delta H-K$} &
    \colhead{$\Delta J-K$} 

  }
  
  \startdata
 RWA 1  & 10.98 & 1.85 & 1.13 & 0.45 & 0.50 & 0.93\\ 
 RWA 2  & 11.90 & 2.74 & 1.78 & 1.19 & 1.36 & 1.64 \\
 RWA 3  & 12.21 & 0.35 & 0.30 & 0.13 & 0.13 & 0.12 \\
 RWA 4  & 10.82 & 1.23 & 0.55 & 1.10 & 0.51 & 0.84 \\
 RWA 5  & 10.78 & 0.67 & 0.48 & 0.19 & 0.22 & 0.40 \\
 RWA 6  & 12.74 & 0.09 & 0.11 & 0.07 & 0.06 & 0.09 \\
 RWA 7  & 12.69 & 0.81 & 0.93 & 0.20 & 0.25 & 0.39 \\
 RWA 8  & 14.40 & 0.43 & 0.32 & 0.47 & 0.44 & 0.52 \\
 RWA 9  & 12.87 & 0.26 & 0.36 & 0.10 & 0.10 & 0.14 \\
 RWA 10 & 13.19 & 0.15 & 0.31 & 0.09 & 0.11 & 0.19 \\
 RWA 11 & 10.96 & 0.14 & 0.18 & 0.07 & 0.12 & 0.12 \\
 RWA 12 & 10.74 & 0.73 & 0.57 & 0.28 & 0.24 & 0.35 \\
 RWA 13 & 10.63 & 0.58 & 0.43 & 0.20 & 0.25 & 0.44 \\
 RWA 14 & 14.07 & 1.49 & 0.73 & 0.42 & 0.70 & 0.99 \\
 RWA 15 & 12.43 & 1.27 & 0.89 & 0.47 & 0.59 & 1.02 \\
 RWA 16 & 11.25 & 0.61 & 0.41 & 0.18 & 0.22 & 0.39 \\
 RWA 17 & 10.38 & 0.94 & 0.51 & 0.37 & 0.25 & 0.59 \\
 RWA 18 & 11.12 & 0.66 & 0.61 & 0.18 & 0.22 & 0.40 \\
 RWA 19 & 10.61 & 0.71 & 0.60 & 0.26 & 0.40 & 0.40 \\
 RWA 20 & 9.76 & 0.87 & 0.89 & 0.17 & 0.15 & 0.20 \\
 RWA 21 & 11.17 & 1.20 & 0.67 & 0.28 & 0.32 & 0.60 \\
 RWA 22 & 11.24 & 0.73 & 0.45 & 0.51 & 0.14 & 0.52 \\
 RWA 23 & 13.48 & 0.48 & 0.35 & 0.12 & 0.25 & 0.34 \\
 RWA 24 & 11.01 & 0.28 & 0.29 & 0.07 & 0.12 & 0.12 \\
 RWA 25 & 10.61 & 0.49 & 0.32 & 0.22 & 0.22 & 0.31 \\
 RWA 26 & 13.71 & 1.33 & 1.05 & 0.49 & 0.35 & 0.71 \\
 RWA 27 & 12.83 & 1.99 & 1.37 & 0.33 & 0.50 & 0.79 \\
 RWA 28 & 15.08 & 0.17 & 0.10 & 0.22 & 0.13 & 0.19 \\
 RWA 28 & 14.24 & 0.17 & 0.23 & 0.16 & 0.10 & 0.25 \\
 RWA 30 & 13.78 & 0.22 & 0.11 & 0.22 & 0.11 & 0.20 \\
 \hline
 Median & ~  & 0.66 & 0.46 & 0.22 & 0.23 & 0.40 \\
 Maximum & ~ & 2.74 & 1.78 & 1.19 & 1.36 & 1.64\\
 Minimun & ~ & 0.09 & 0.10 & 0.07 & 0.06 & 0.09\\ 

\enddata
\end{deluxetable}

\end{document}